\documentclass[%
 twocolumn,
 bibnotes,
 amsmath,amssymb,
 aps, physrev,
 prstab,
]{revtex4-2}

\usepackage{graphicx}
\usepackage{dcolumn}
\usepackage{bm}
\usepackage{subcaption}
\usepackage{caption}
\usepackage{ragged2e} 
\usepackage{array, booktabs, multirow, threeparttable}
\usepackage{color,soul}
\usepackage{xcolor}
\usepackage{listings}
\usepackage{placeins}

\begin{document}

\preprint{APS/123-QED}

\title{\textbf{Short-Pulse Driven Radiofrequency X-Band Photoinjector: \\
Electromagnetic Properties and Beam Dynamics in the Transient Regime}}

\author{Gongxiaohui Chen}
 \email{Contact author: b288079@anl.gov}
 \affiliation{Argonne National Laboratory, 9700, S. Cass Avenue, Lemont IL 60539, USA}
 
\author{Chunguang Jing}
\affiliation{Euclid Techlabs LLC, 367 Remington Blvd, Bolingbrook, IL 60440, USA}
 \affiliation{Argonne National Laboratory, 9700, S. Cass Avenue, Lemont IL 60539, USA}

\author{Philippe Piot}
 \affiliation{Argonne National Laboratory, 9700, S. Cass Avenue, Lemont IL 60539, USA}

\author{John Power}
 \affiliation{Argonne National Laboratory, 9700, S. Cass Avenue, Lemont IL 60539, USA}

\date{\today}

\begin{abstract}

This paper presents a study of the radiofrequency (RF) characteristics and beam dynamics of an X-band photogun (Xgun) operating in the transient state. The photoinjector is designed to operate with short RF pulses (9~ns) to achieve high accelerating gradients. Short-pulse operation potentially reduces breakdown risks, as experimentally demonstrated by achieving a gradient exceeding 350~MV/m. However, the short pulse duration causes the cavity to operate in a transient regime where the electromagnetic field deviates from the conventional steady-state condition. To investigate the effects of deviations from steady-state operation, the time-dependent spatial evolution of the cavity fields was examined using both 9-ns and 50-ns RF pulses. The 50-ns pulse served as a reference to characterize the cavity behavior under a fully filled steady-state condition. Beam dynamics simulations were conducted to explore the impact of transient RF effects on beam kinetic energy, transverse emittance, and bunch length; these simulations employed a new dynamic field mapping approach to model transient RF fields. 

\end{abstract}

\maketitle


\section{Introduction}
The development of high-brightness photoinjectors is crucial for advancing a wide range of scientific applications, including next-generation free-electron lasers (FELs)\cite{emma2010first,piot:fls2023-mo4c2}, compact X-ray sources\cite{graves2014compact}, and ultrafast electron diffraction or microscopy (UED/UEM)\cite{qi2020breaking, li2014single}. To generate the low emittance, high charge beams required for these applications, photoinjectors must integrate two key features ($i$) cathodes with low thermal emittance, which sets the lowest achievable emittance, and ($ii$) high fields at the cathode to mitigate space-charge effects, thus preserving the low-emittance electron bunch immediately after the initial emission process. The transverse brightness ($\mathcal{B}$) of the beam, which scales with the accelerating gradient ($E$) at the cathode surface as $\mathcal{B}\propto E^m/\textit{MTE}$, demonstrates the critical relationship between $\mathcal{B}$ and $E$. Here, $m$ depends on the beam's transverse-to-longitudinal aspect ratio, and $\textit{MTE}$ represents the mean transverse energy of the cathode\cite{filippetto2014maximum}. Efforts to enhance brightness have thus focused on both novel cathode materials to reduce $\textit{MTE}$ and on applying high electric fields to photocathodes, which have driven the development of RF photoinjectors—critical components for FEL and other high-brightness applications. Standard RF photoinjectors routinely operate with photocathode surface fields around 120~MV/m. Efforts at the Argonne Wakefield Accelerator (AWA) facility aim to attain higher surface fields.

The primary challenge in achieving higher surface and accelerating fields in normal-conducting accelerating structures is the potential risk of destructive RF breakdowns, which can damage the structure and limit performance. The breakdown rate ($\textit{BDR}$) has been empirically related to both the accelerating field ($E$) and the RF pulse duration ($\tau$) by the scaling law~\cite{grudiev2009new}, 
\begin{eqnarray}
    \textit{BDR} \propto E^{30}\tau^5. 
\end{eqnarray}
Recent breakthroughs at the AWA have demonstrated that short RF pulses ($\sim$10~ns) can suppress breakdown initiation, enabling gradients exceeding 400~MV/m in an X-band photogun~\cite{tan2022demonstration,chen2025design} and exceeding 300~MV/m in a single-cell accelerating structure~\cite{shao2022demonstration}. However, due to the short RF-pulse duration, the structure may not reach its steady-state operation. This transient state condition generally yields a different electromagnetic (EM) field distribution compared to the steady-state potentially affecting the beam dynamics. This could be especially problematic in an RF photoinjector, where the beam dynamics are strongly dependent on the field topology given the beam's low energy.

In this work, we first present an overview of the two-beam acceleration (TBA) based short-pulse RF generation system, which produces 9-ns RF pulses at 11.7~GHz. The TBA technique was proposed four decades ago~\cite{hopkin-1984-a,schnell-1985-a,sessler:1987-a} and is critical to producing short RF pulses. Then, we use the X-band photogun (Xgun) as a prototype, which is designed for short RF pulse (9~ns) operation at high-gradients. Unlike long-pulse powered systems that reach steady-state conditions, the Xgun remains under-filled, operating entirely in a transient state throughout the RF pulse duration. In this regime, the EM fields behave differently from those in a steady state, and their time-dependent evolution can also affect beam properties. To investigate these effects, we analyze the spatial and temporal evolution of cavity fields using both 9-ns and 50-ns pulses, with the longer 50-ns pulse twice the duration required to reach steady-state conditions, serving as a steady-state reference. Additionally, preliminary beam dynamics simulations were conducted using a dynamic field mapping approach to quantify the impact of transient RF fields on beam performance.

\section{Short RF Pulse Generation}

\begin{figure}[h]
    \centering
    \includegraphics[width=\linewidth]{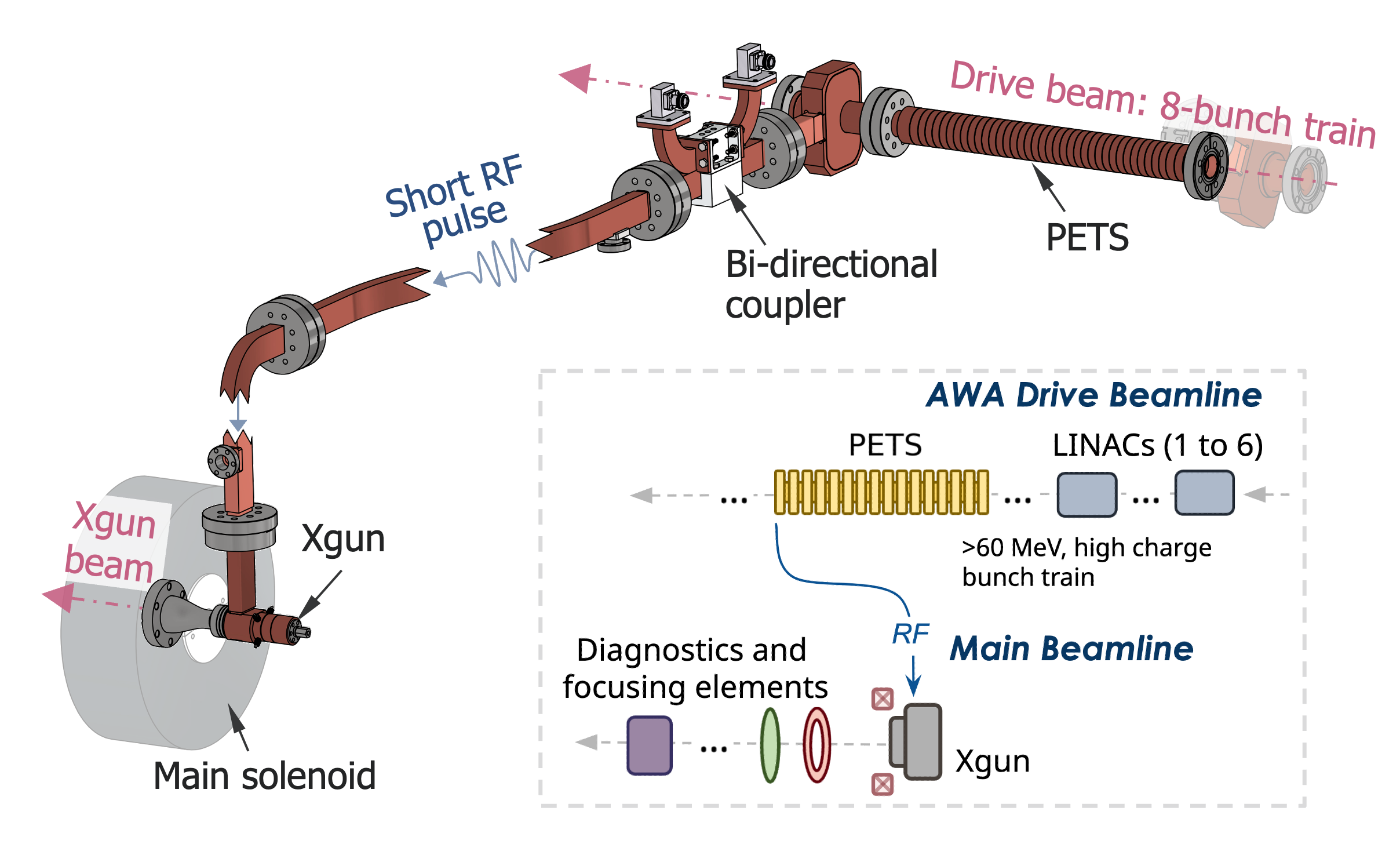}  
    \caption{\justifying Schematic diagram of the TBA system, including PETS from the drive beamline and the Xgun from the main beamline. The RF source generated by the PETS is then transferred to the Xgun. Inset: block diagram of the TBA system.}
    \label{fig:PETS}
    \end{figure}

At the AWA facility, two-beam acceleration (TBA) technology is employed to generate short RF pulses. The practical realization of TBA-technique was pioneered at CERN's Compact LInac Collider (CLIC) test facility~\cite{corsini-2017-a}. In this process, a high-charge drive beam passes through a power extraction and transfer structure (PETS) to produce high-power RF source, which is then transferred to the main beamline (here, the Xgun beamline) to power the Xgun cavity. A diagram of AWA's TBA system appears in Fig.~\ref{fig:PETS}. The drive beam is composed of eight bunches, each separated by 769~ps, and is accelerated by 8 consecutive buckets of the AWA linear accelerator, which operates at a frequency of 1.3~GHz. In its current configuration, a metallic corrugated PETS with fundamental-mode frequency at 11.7~GHz (corresponding to the $9^{\mbox{\small th}}$ harmonic of 1.3~GHz) is employed; more details on the PETS design and performance are available in Ref.~\cite{jing2012high,shi2013development,jing2018electron}. 

\begin{figure}[t]
    \centering
    \includegraphics[width=0.95\linewidth]{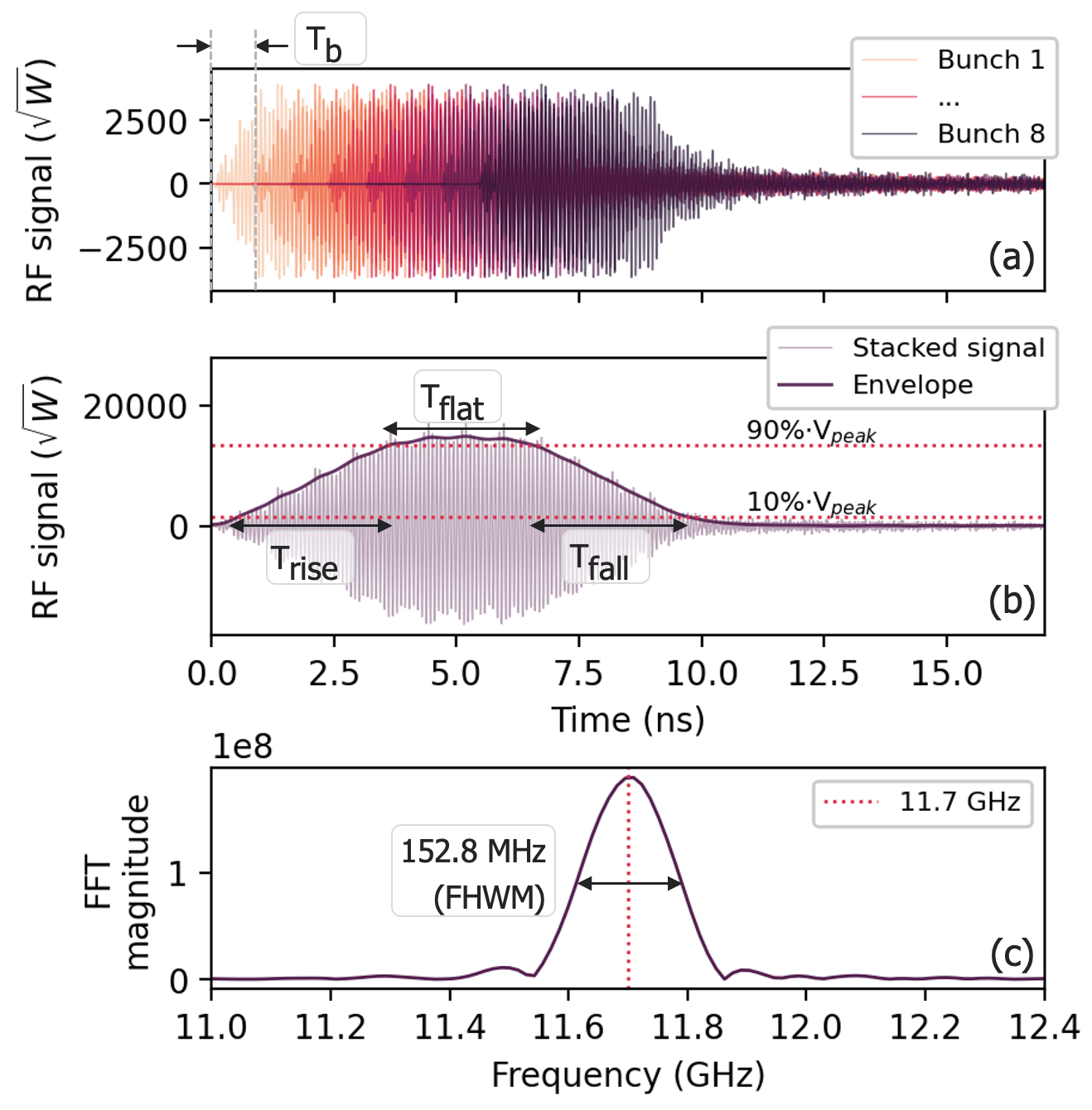} 
    \caption{\justifying (a) Simulated RF signal for a single bunch, with signals sequentially shifted by $T_\text{b}=0.769$~ns to represent the 8-bunch train. (b) Stacked RF signal from the 8-bunch train shown in (a) and its envelope, highlighting $T_\text{rise}$, $T_\text{flat}$, and $T_\text{fall}$. (c) FFT spectrum of the RF signal shown in (b) with the center frequency and FWHM bandwidth of 11.7~GHz and 152.8~MHz respectively.}
    \label{fig:PETS_rf}
    \end{figure}

The output RF signal from the PETS was simulated using the CST~\cite{CST} wakefield solver. In the simulation, a single 25-nC bunch following a longitudinal Gaussian distribution with an rms bunch length of $\sigma_z=1.5$~mm was considered. The output RF pulse produced by this single bunch is shown in Fig.~\ref{fig:PETS_rf}~(a). The pulse has a duration given by the length of the PETS structure and group velocity of the excited fundamental mode. To construct the RF pulse associated with the train of 8 bunches, the single-bunch RF pulse is sequentially shifted by $T_\text{b}=769$~ps. The RF signal produced from the 8-bunch train is shown in Fig.~\ref{fig:PETS_rf}~(b). The pulse follows a ``trapezoidal" shape with peak amplitude close to 8 times the single-bunch RF pulse, showcasing the intra-bunch build-up and coherent enhancement. The rise time, defined as the time for the envelope to increase from 10\% to 90\% of its peak value, is $T_\text{rise}=3.08$~ns; the flat-top duration, where the envelope remains within $\pm$10\% of its peak value, is $T_\text{flat}=2.97$~ns; the fall time, from 90\% to 10\%, is $T_\text{fall}= 3.17$~ns. Figure~\ref{fig:PETS_rf}~(c) shows the fast Fourier transform (FFT) spectrum of the RF pulse produced by the bunch train, confirming the center frequency of 11.7~GHz and a full-width at half-maximum (FWHM) bandwidth of 152.8~MHz. 

\begin{figure}[h]
    \centering
    \includegraphics[width=\linewidth]{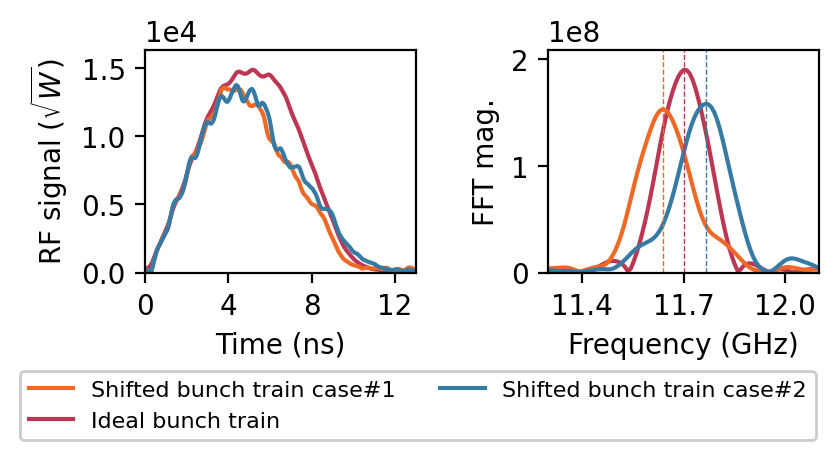}  
    \caption{\justifying Comparison of RF signals generated by three different bunch-train configurations: an ideal 8-bunch train; Case$\#$1 introduces a +1° phase shift (+2~ps) on all delay stages, and Case$\#$2 with -1° phase shift (-2~ps). Left: RF envelopes. Right: FFT spectra.}
    \label{fig:shifted_bunch}
\end{figure}

\begin{figure}[t]
    \centering
    \includegraphics[width=\linewidth]{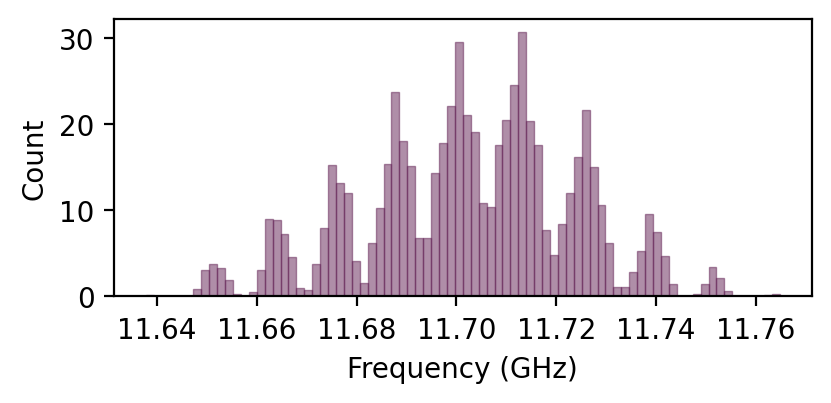}  
    \caption{\justifying Histogram of center frequencies obtained from scanning all three delay stages, with each stage varied from –1° to 1° in 0.05° increments, based on 68921 calculated combinations.}
    \label{fig:PETS_freq}
\end{figure}

Additionally, the center frequency of the RF signal generated by the PETS can be actively tuned by adjusting the temporal structure of the electron bunch train,  enabling compensation for small frequency mismatches or adaptation to specific operational requirements, without significant impact on the peak RF power. This is achieved through the AWA laser splitter system (Appendix~\ref{beam_splitter}), where three delay stages control the temporal delay $T_\text{b}$ between the bunches. By introducing calibrated phase shifts to the delay stages, the cumulative timing of the bunches is reconfigured, directly modifying the RF output frequency. For example, two cases of $\pm$1° (L-band) phase variation ($\pm$2~ps) were analyzed: Case$\#$1, a +1° phase shift was applied at each of the three delay stages, resulting in a cumulative phase shift pattern across the bunch train of (0°, 1°, 1°, 2°, 1°, 2°, 2°, 3°); Case$\#$2, a -1° phase shift on all three delay stages resulting a phase shift pattern of (0°, -1°, -1°, -2°, -1°, -2°, -2°, -3°). As shown in Fig.~\ref{fig:shifted_bunch}, these bunch spacing variations caused the center frequency to adjust from the designed value of 11.7~GHz, to 11.64~GHz and 11.76~GHz for case$\#$1 and case$\#$2, respectively. Figure~\ref{fig:PETS_freq} further quantifies the tunable range, showing a span of $\pm 0.06$~GHz, given the scanning range of $\pm 1$° in 0.05° increments for each delay stage. Extensive experimental RF data from the PETS can be found in Ref \cite{jing2012high, peng2019generation}. For simplicity, in the following studies, an ideal trapezoidal RF pulse with T$_\text{rise}$=T$_\text{fall}$=3~ns and a flat top with adjustable duration (T$_\text{flat}$) was adopted, as shown in Fig.~\ref{fig:E_cathode} (top). The frequency of the RF pulse was fixed to the value optimal for Xgun operation. 

Though a klystron could be considered as an alternative power source and specifically designed for optimal Xgun performance, current pulse compression technology cannot provide phase-stable, sub-10~ns, sub-GW RF power. This limitation positions the TBA technique as the only viable solution for meeting the high-power and short-pulse requirements of the Xgun.

\section{RF Characteristics in the Transient State}
\subsection{Overview of the Xgun Design}

The Xgun is a 1.5-cell, strongly over-coupled standing-wave (SW) RF cavity specifically designed for short (9~ns) RF pulse operations. In the context of short-pulse operation, having a rapid RF fill-time ($t_{\text{fill}}$) is critical for reaching the desired accelerating gradient within the limited RF pulse duration. For an SW structure, the fill-time is defined by
\begin{equation}
\begin{aligned}
t_{\text{fill}}={2Q_{\text{load}}}/{\omega}, 
\end{aligned}
\end{equation}
where $Q_{load}$ is the loaded $Q$ factor of the Xgun and $\omega$ its fundamental-mode operating angular frequency. At a given operating frequency, achieving a fast fill-time requires a low $Q_{load}$. In the current Xgun design, $Q_{load}=182$, resulting in $t_{\text{fill}}\sim$5~ns, which corresponds to reaching 63\% of the steady-state field-amplitude level. Since the field amplitude build-up in the cavity is described by
\begin{equation}
E(t)=E_{\text{max}}(1-e^{-t/t_\text{{fill}}}),
\end{equation}
where $E_{\text{max}}$ represents the steady-state field amplitude. 
Numerically, given the $t_{\text{fill}}\simeq 5$~ns, reaching the steady-state (99\% amplitude level) requires about 23~ns. Thus, the Xgun operates in a transient state throughout the entire RF pulse duration (i.e. 9~ns). In the following sections, the RF characteristics and beam dynamics in the transient state of the Xgun are discussed in detail.

\begin{figure}[b]
    \centering
    \includegraphics[width=0.7\linewidth]{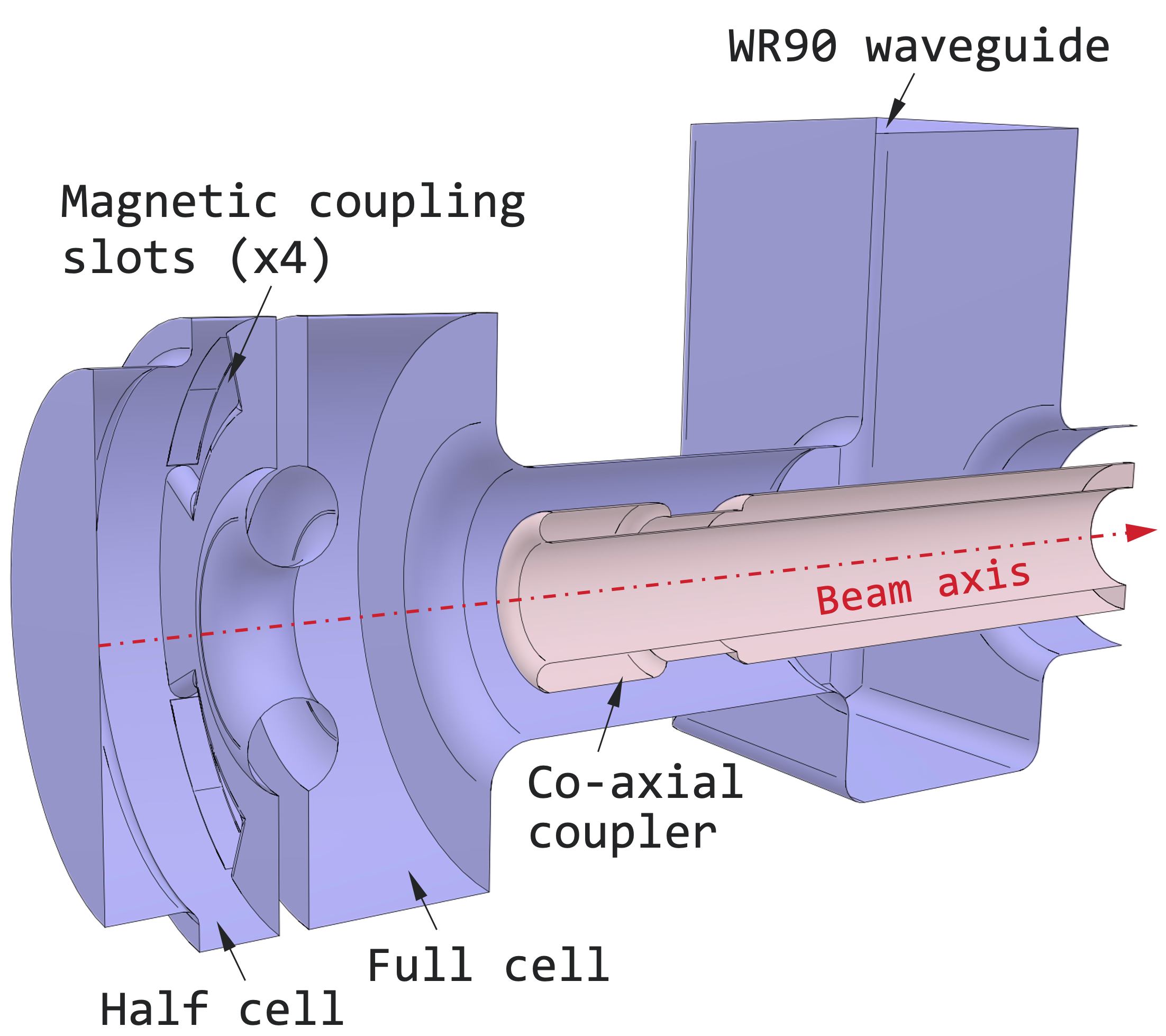}
    \caption{\justifying Mid-plane view of the Xgun's vacuum model (shaded blue volume). The shaded pink volume assembly corresponds to the coaxial-coupler inner conductor, which also serves as the beam channel.}
\label{fig:Xgun_design}
\end{figure}

\begin{figure}[t]
    \centering
    \includegraphics[width=0.95\linewidth]{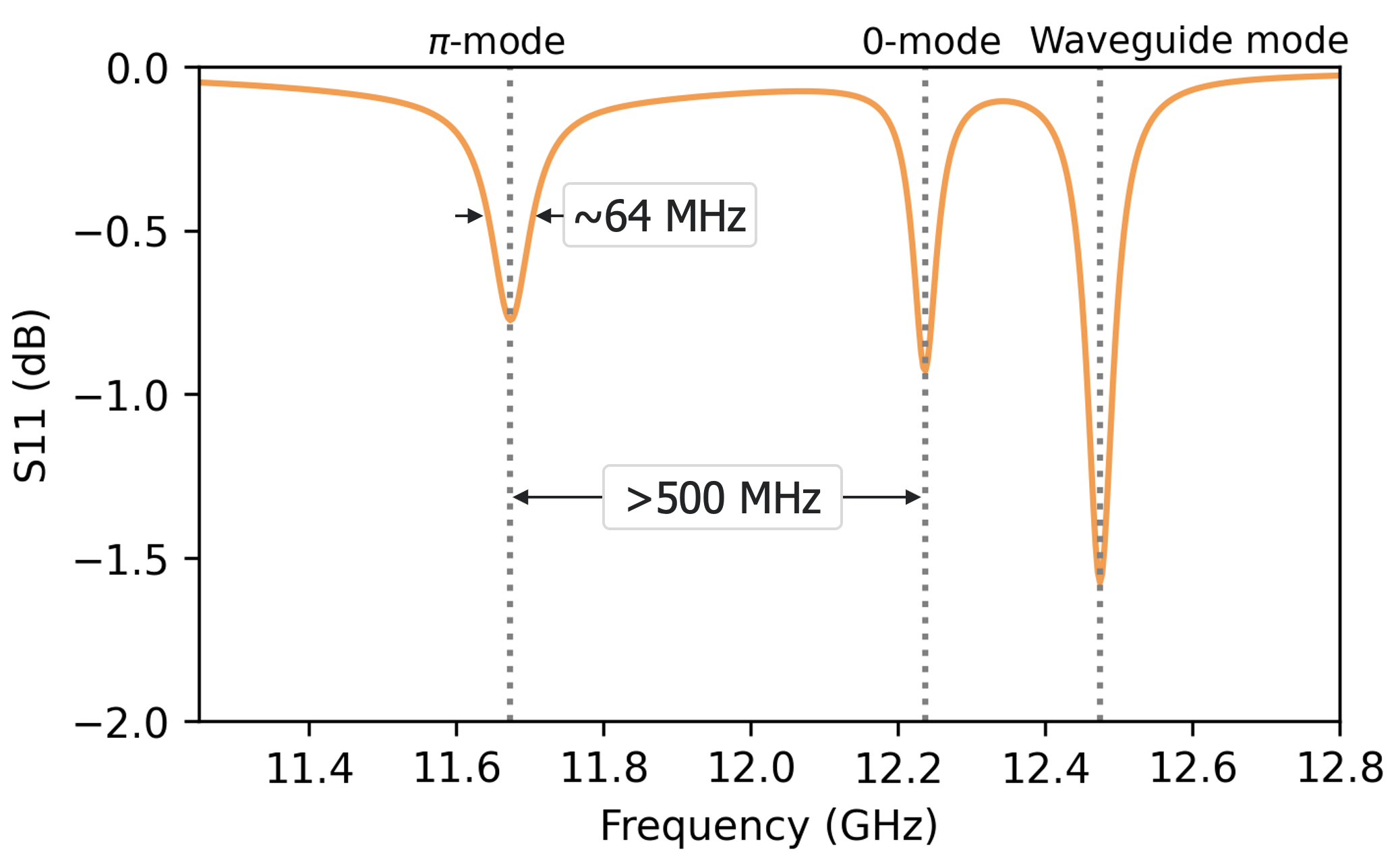}
    \caption{\justifying The simulated $S_{11}$ reflection scattering parameter for the Xgun.}
    \label{fig:S11}
\end{figure}

Figure~\ref{fig:Xgun_design} presents a cut-away view of the Xgun cavity. The design incorporates four magnetic coupling slots between the half and full cells to improve the coupling strength between the cells while also increasing the mode separation. Given that short RF pulses inherently have a broad bandwidth, enlarging the mode separation ensures that only the desired $\pi$-mode is excited. Figure~\ref{fig:S11} gives the \textit{S}$_{11}$ reflection scattering parameter simulated using CST indicating a mode separation $> 500$~MHz, larger than the RF-pulse bandwidth of 152.8~MHz. The basic RF parameters are listed in Table~\ref{table:rf_param}, and additional details on the Xgun design can be found in Ref.~\cite{chen2025design}.

\begin{table}[h]
    \caption{\justifying Basic RF parameters of the Xgun.}
       \begin{ruledtabular}
       \begin{tabular}{lc}
           \textbf{Parameter}                              & \textbf{Value} \\
           \hline
            Frequency                                      & 11.68~GHz        \\ 
            $Q_\text{0}$                                   & 4256             \\ 
            $Q_\text{ext}$                                 & 191              \\ 
            $Q_\text{load}$                                & 182              \\ 
            Coupling coefficient ($\beta$)                 & 22.3             \\
            $P_\text{input}$                               & 136~MW           \\
            RF pulse duration ($\tau$)                     & 9~ns             \\
            E field on cathode ($E_\text{cath}$)\footnote{This value was calculated using the CST time-domain solver with a 9-ns RF pulse shown in Fig.~\ref{fig:E_cathode}.}  & 350~MV/m         
       \end{tabular}
       \end{ruledtabular}
       \label{table:rf_param}
    \end{table}

\subsection{RF characteristics of the Xgun}

This section describes the RF characteristics during this transient state are investigated, with a focus on the field build-up process and the spatial and temporal evolution of the cavity fields throughout the pulse duration.

\subsubsection{Frequency-dependent $E_z$}

As discussed in the previous section, to enable short-pulse operation, the $Q_\text{load}$ is lowered to allow for a rapid fill-time of the cavity. This design choice inherently results in a relatively wide resonance bandwidth around the operating frequency (as $\Delta f \propto 1/Q_\text{load}$), as shown in Fig.~\ref{fig:S11}. Thus, the cavity becomes more sensitive to slight variations in the input RF frequency, which can affect the field performance.

\begin{figure}[h]
    \centering
    \includegraphics[width=0.95\linewidth]{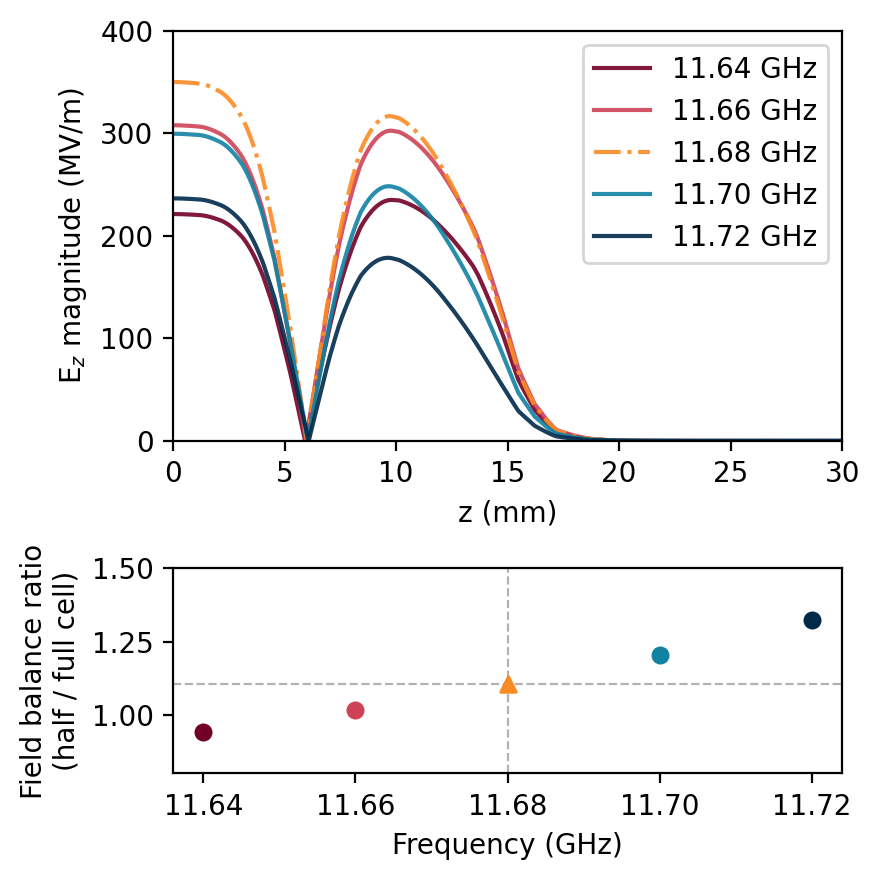}  
    \caption{\justifying Top: simulated on-axis $E_z$ field distribution at different frequencies using the CST F-domain solver. Bottom: field balance ratio between the half-cell and the full-cell.}
    \label{fig:Ez_F_domain}
\end{figure}

Figure~\ref{fig:Ez_F_domain} presents the simulated on-axis $E_z$ field distribution obtained using the CST frequency-domain (F-domain) solver, with the field monitors set to different frequencies (from 11.64~GHz to 11.72~GHz) within the $\pi$-mode bandwidth. The simulated $E_z$ from the F-domain solver represents the EM fields at steady-state. For consistent comparison, all $E_z$ fields were scaled based on the same input power, such that the global highest field amplitude on the photocathode is 350~MV/m. Given the same input power, the 11.68~GHz frequency produces the highest field, albeit at the expense of an imperfect field balance between the half-cell and the full-cell compared to the ideally-balanced field distribution at 11.66~GHz. Generating a high field at the photocathode can be advantageous from a beam dynamics perspective, as this increased field strength can improve beam quality.

At the current proof-of-concept stage, our primary objective is to achieve a high-gradient field on the cathode and demonstrate its effectiveness in enhancing beam brightness through initial high-gradient acceleration. Therefore, a frequency of 11.68~GHz was selected. In addition, we assume an RF pulse generated from a train of 8 bunches traversing the PETS to follow a trapezoidal envelope for the time-domain simulations presented in the remainder of this paper.

\subsubsection{RF build-up in the Xgun}

\begin{figure}[t]
    \centering
    \includegraphics[width=0.95\linewidth]{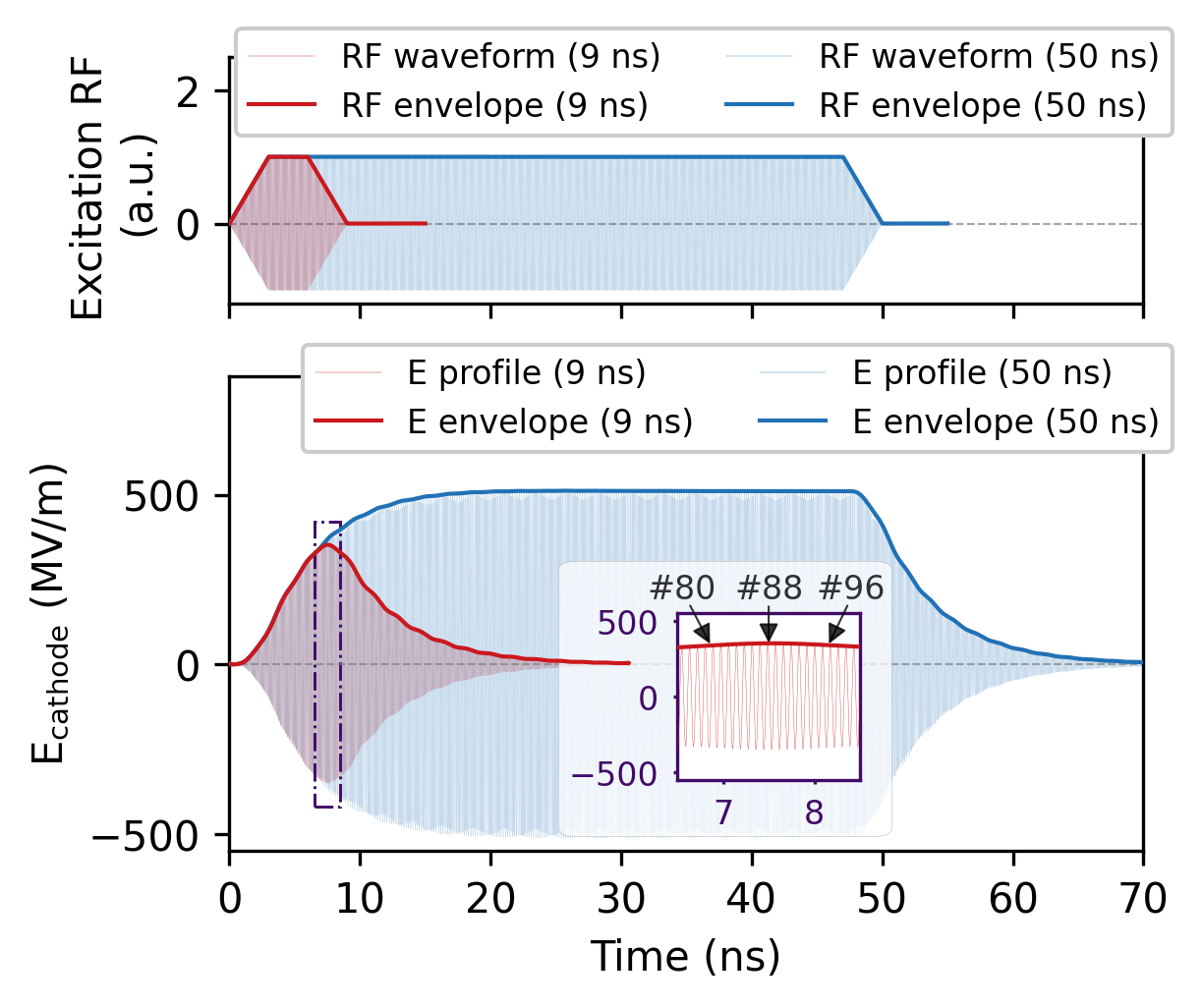}
    \caption{\justifying Top: trapezoidal-excitation RF pulses [9 (blue) and 50 (red) ns] used in the CST time-domain simulations, with the frequency of 11.68~GHz. Bottom: evolution of the electric-field amplitude on the photocathode surface. The inset in the lower plot is a zoomed-in view for $t\in [ 6.5, 8.5]$~ns, i.e., in the temporal window where the electric field reaches it peak value. The $\#$80, $\#$88, and $\#$95 labels indicate the RF cycles numbers since $t=0$.}
    \label{fig:E_cathode}
\end{figure}

To analyze the transient field characteristics and the build-up process within the cavity, CST's time-domain solver was used. To compare transient and steady-state conditions, two RF excitation signals were employed, as shown in Fig.~\ref{fig:E_cathode} (top). The first case is a short 9-ns full-width trapezoidal RF pulse, an idealized model of a PETS-generated RF signal with parameters $T_\text{rise}=T_\text{flat}=T_\text{fall}=3$~ns. The second case is a 50-ns full-width RF pulse with the same $T_\text{rise}=T_\text{fall}=3$~ns, and $T_\text{flat}=44$~ns, corresponding to the steady-state condition. The frequency of the RF pulse was fixed at 11.68~GHz for both cases.

As a first step toward gaining insight into the field build-up on the cathode surface, a field probe was placed on the cathode surface in CST to monitor the time-dependent evolution of the electric field, as shown in Fig.~\ref{fig:E_cathode} (bottom). For both cases, the input power was fixed at 136~MW. Under short-pulse operation with the 9-ns pulse, this power drives the Xgun to achieve a peak field of 350~MV/m on the cathode surface. When driven by the ``long" 50-ns RF pulse, the Xgun is fully filled and reaches steady-state conditions around 23~ns, achieving a peak gradient of $> 500$~MV/m. In contrast, with the 9-ns RF pulse, the Xgun remains in the transient state during the entire pulse duration. As illustrated in the zoomed-in inset of Fig.~\ref{fig:E_cathode} (bottom), the field on cathode reaches its peak value at 7.489~ns, corresponding to the 88$^\text{th}$ RF cycle at the nominal frequency of 11.68~GHz. Additionally, the cathode field remains above 330~MV/m for an extended period between 6.8 and 8.2~ns, covering $>16$ RF cycles. This extended duration of high-field provides a broad operational window for high-gradient beam tests, despite operating in a transient regime.

Having characterized the temporal evolution of the field at a specific location on the photocathode surface, the next step is to understand the spatial evolution of the field through the cavity. For a SW cavity, the on-axis $E_z(z,t)$ is expressed as:
\begin{equation}
    E_z(z,t)=E_{\text{z,0}}(z) \text{cos}(2\pi f t),
\label{eq:standing_wave}
\end{equation}
where $E_{\text{z,0}}(z)$ represents the field profile on-axis, $f$ is the frequency, and $t$ is the time. This expression represents the steady-state behavior of a SW cavity, where the field oscillates sinusoidally over time. However, the short-pulse operation keeps the cavity in a transient state, and the field $E(z,t)$ dynamically evolves as the RF power gradually fills the cavity, reaching its peak before beginning to decay. This transient behavior is critical for understanding the cavity's performance, as it influences the timing and efficiency of energy transfer, as well as the stability and quality of the beam produced. It is expected that in the transient regime, the field evolution is not accurately described by Eq.~\ref{eq:standing_wave}.

\begin{figure*}[htbp!]
    \centering
    \begin{minipage}{.48\linewidth}
        \begin{subfigure}[t]{.9\linewidth}
            \includegraphics[width=\textwidth]{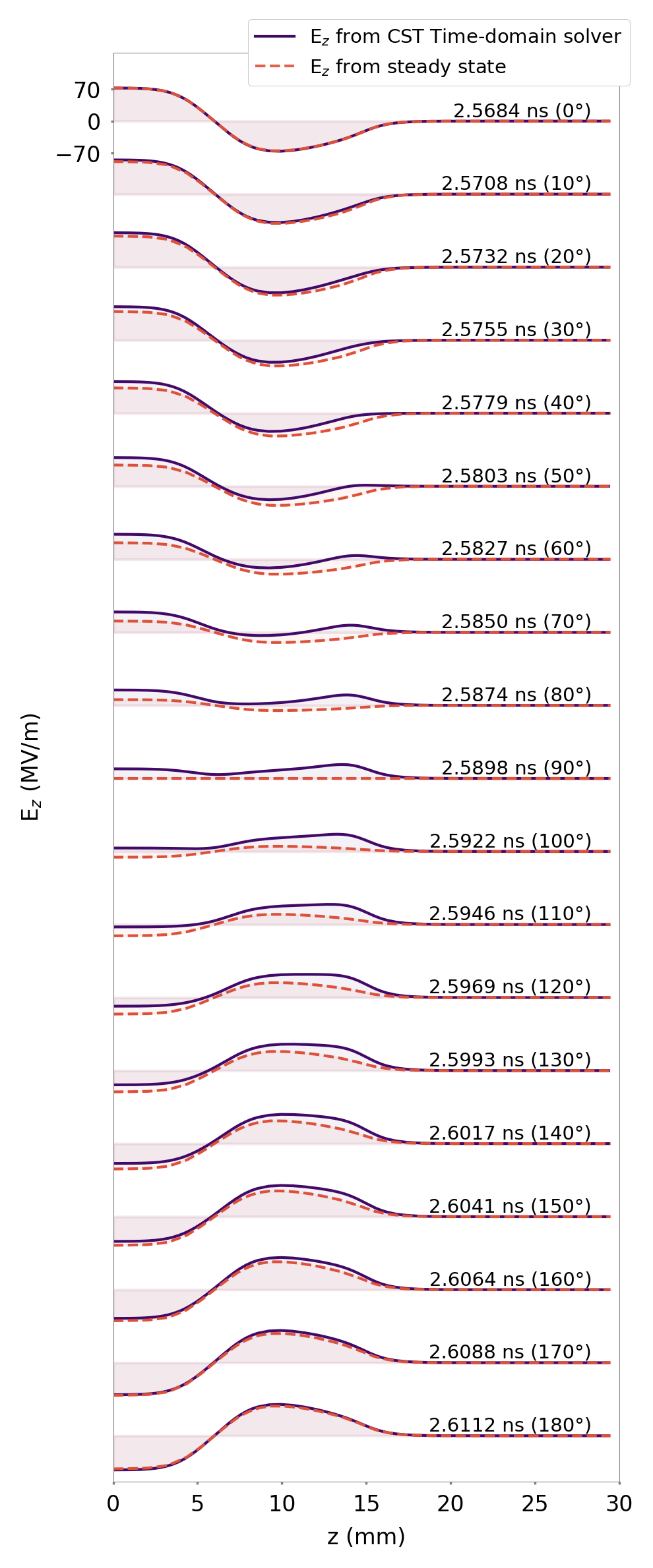}
            \caption{RF bucket No. 30}
            \label{fig:rf_bucket_30}
        \end{subfigure}
    \end{minipage}
    \hfill
    \begin{minipage}{.48\linewidth}
        \begin{subfigure}[t]{.9\linewidth}
            \includegraphics[width=\textwidth]{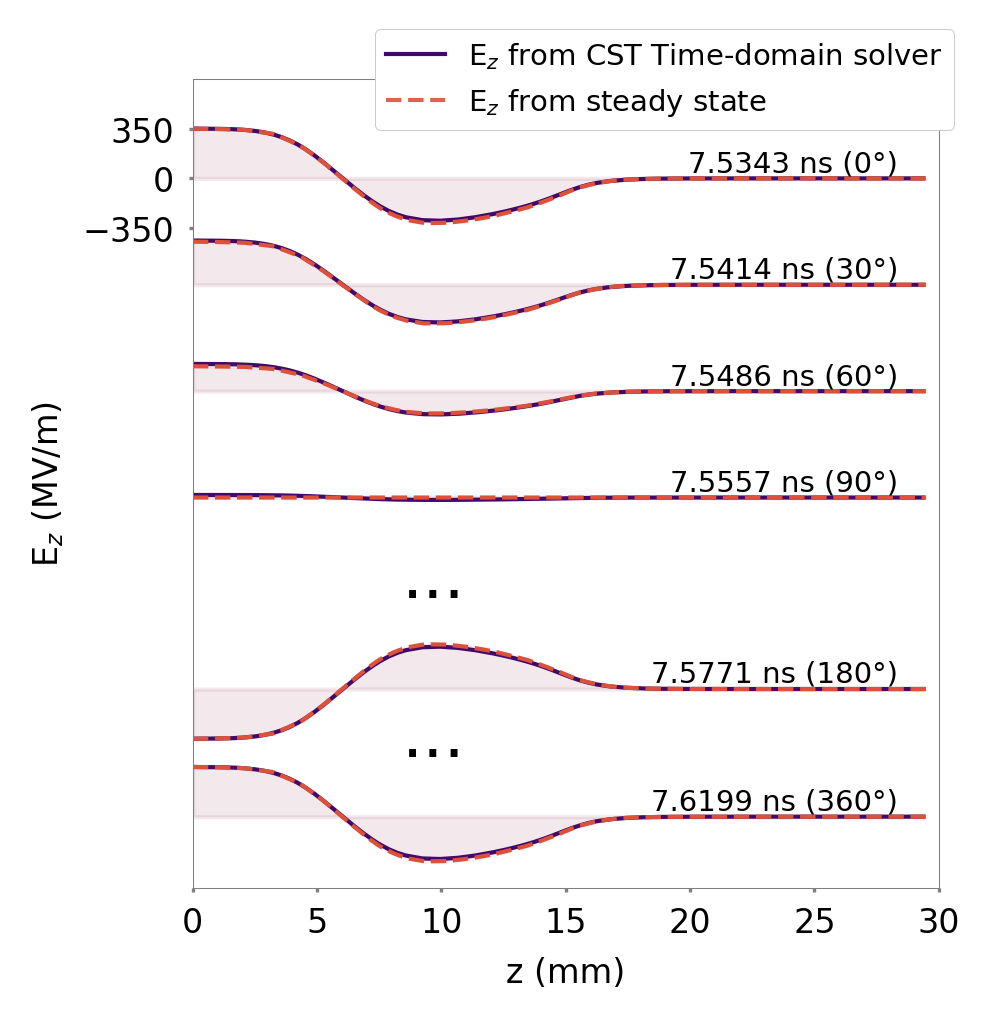}
            \caption{RF bucket No. 88}
            \label{fig:rf_bucket_88}
        \end{subfigure} \\
        \begin{subfigure}[b]{.9\linewidth}
            \includegraphics[width=\textwidth]{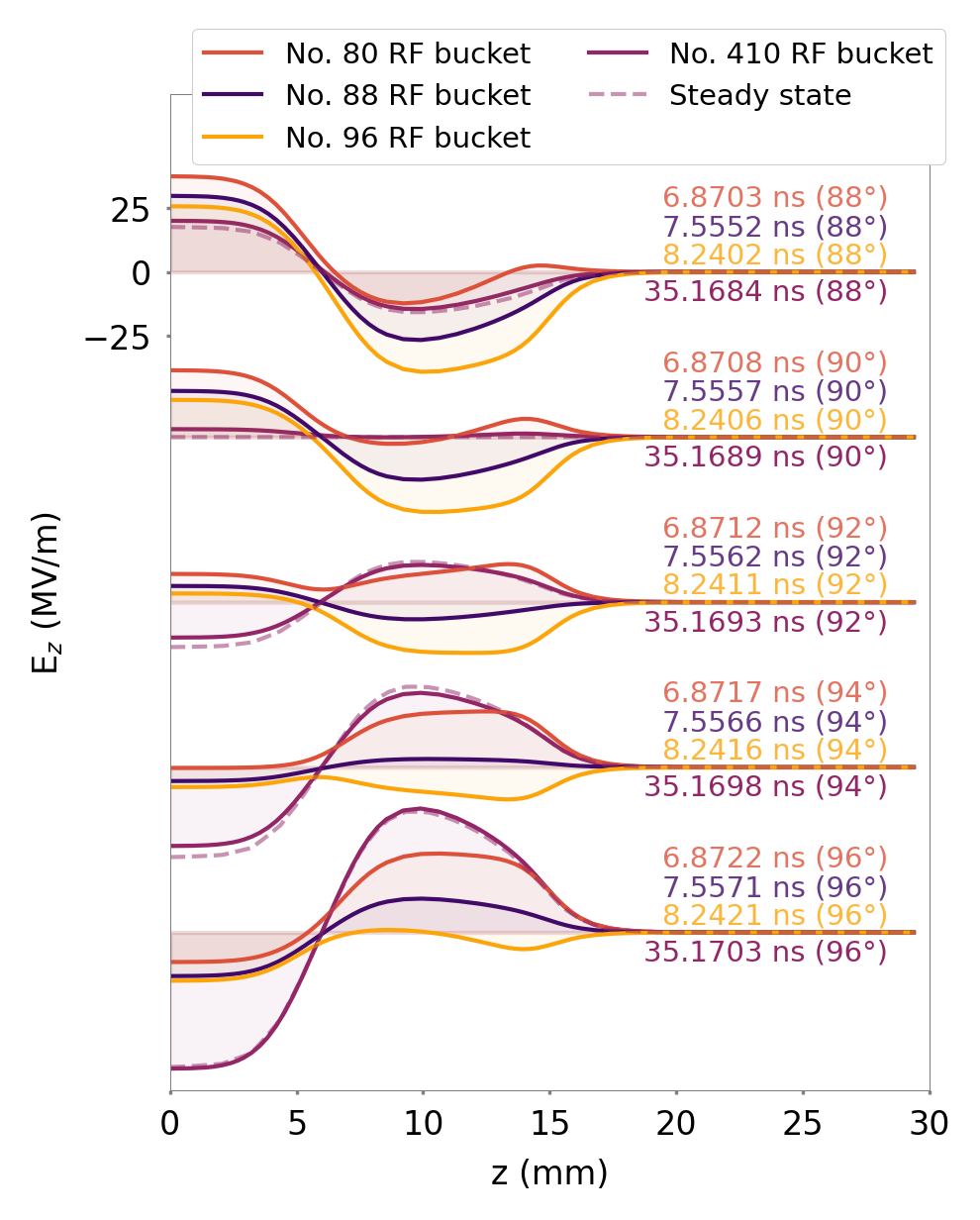}
            \caption{Comparison of RF buckets (\#80, \#88, \#96, \#410)}
            \label{fig:rf_bucket_comp}
        \end{subfigure}
    \end{minipage}
    \caption{\justifying Temporal snapshots of the spatial distribution of $E_z(z,t)$ along the Xgun $z$-axis. Solid lines: simulated field-distributions from the CST T-domain solver. Dashed lines: steady-state field-distribution from the F-domain solver at 11.68~GHz, with the magnitude scaled to match the peak field of the specific RF bucket. (a) $E_z$ field evolution in the 30$^\text{th}$ RF bucket (starting from 2.5684~ns) plotted for every 10° phase interval. (b) $E_z$ field evolution in the 88$^\text{th}$ RF bucket (starting from 7.5343~ns), which corresponds to the bucket with the peak field on the cathode surface. (c) Comparison of $E_z$ fields from different buckets (\#80, \#88, \#96, and \#410), with bucket \#410 representing the simulated results using the long 50-ns rf pulse, corresponding to the fully-filled cavity at steady-state conditions. For comparison, the dashed lines in (c) are the steady-state $E_z$ scaled to match the \#410 bucket peak field.}
    \label{fig:Ez_in_diff_cycles}
\end{figure*}

\begin{figure*}[t]
    \begin{subfigure}[b]{0.95\linewidth}
        \includegraphics[width=\textwidth]{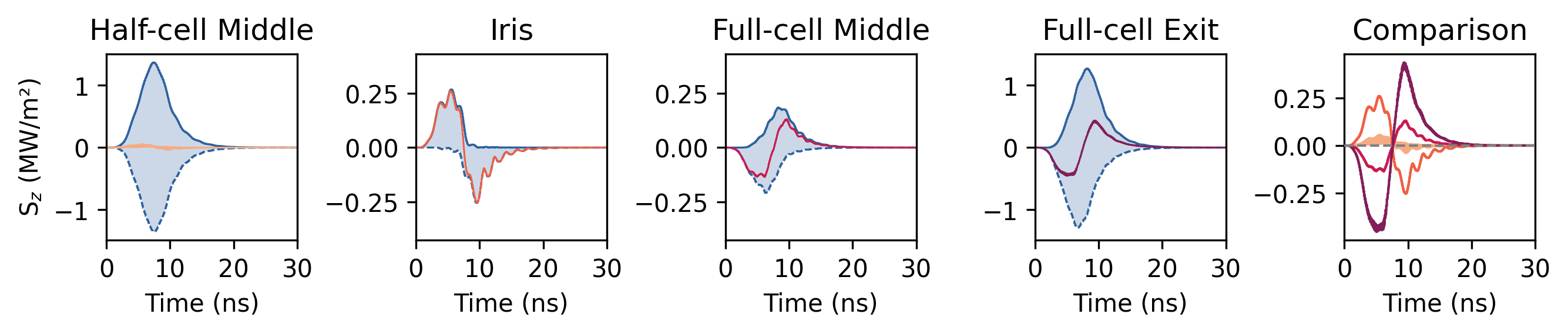}
        \caption{$S_z(t)$ of the 9~ns RF pulse-driven Xgun.}
        \label{fig:9ns_Sz_total}
    \end{subfigure}
    \hfill
    \begin{subfigure}[b]{0.95\linewidth}
        \includegraphics[width=\textwidth]{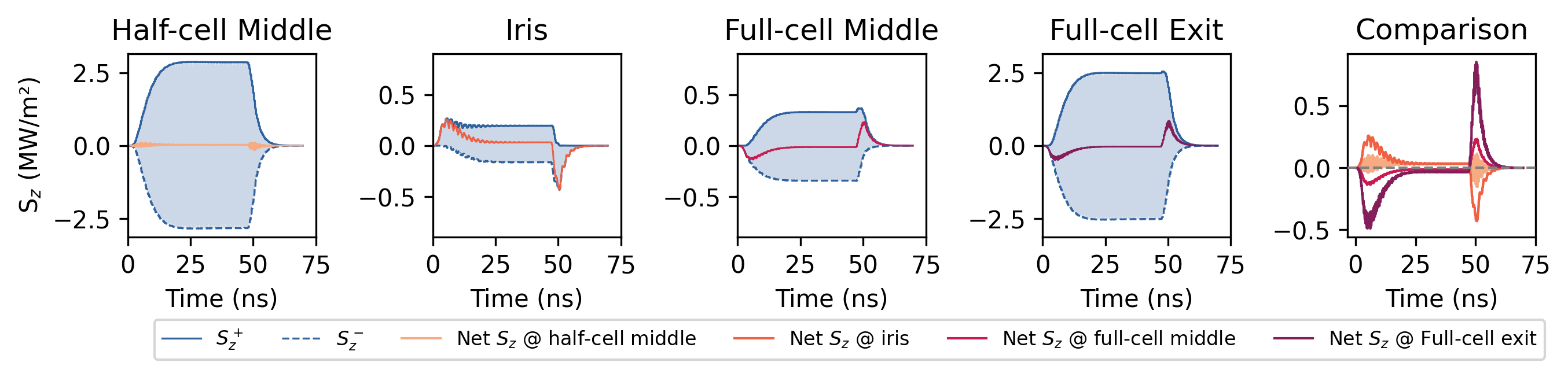}
        \caption{$S_z(t)$ of the 50~ns RF pulse-driven Xgun.}
        \label{fig:50ns_Sz_total}
    \end{subfigure}
\caption{\justifying Poynting vector ($S_z$) as a function of time at different locations of the Xgun as described in the subtitles. Subfigures in the last column compare the $S_z$ at all locations. The gray dashed line indicates zero power flow (where $S_z = 0$). (a) 9~ns RF pulse powered Xgun. (b) 50~ns RF pulse powered Xgun.}
\label{fig:Sz_total}
\end{figure*}

\begin{figure}[b!]
    \centering
    \includegraphics[width=0.95\linewidth]{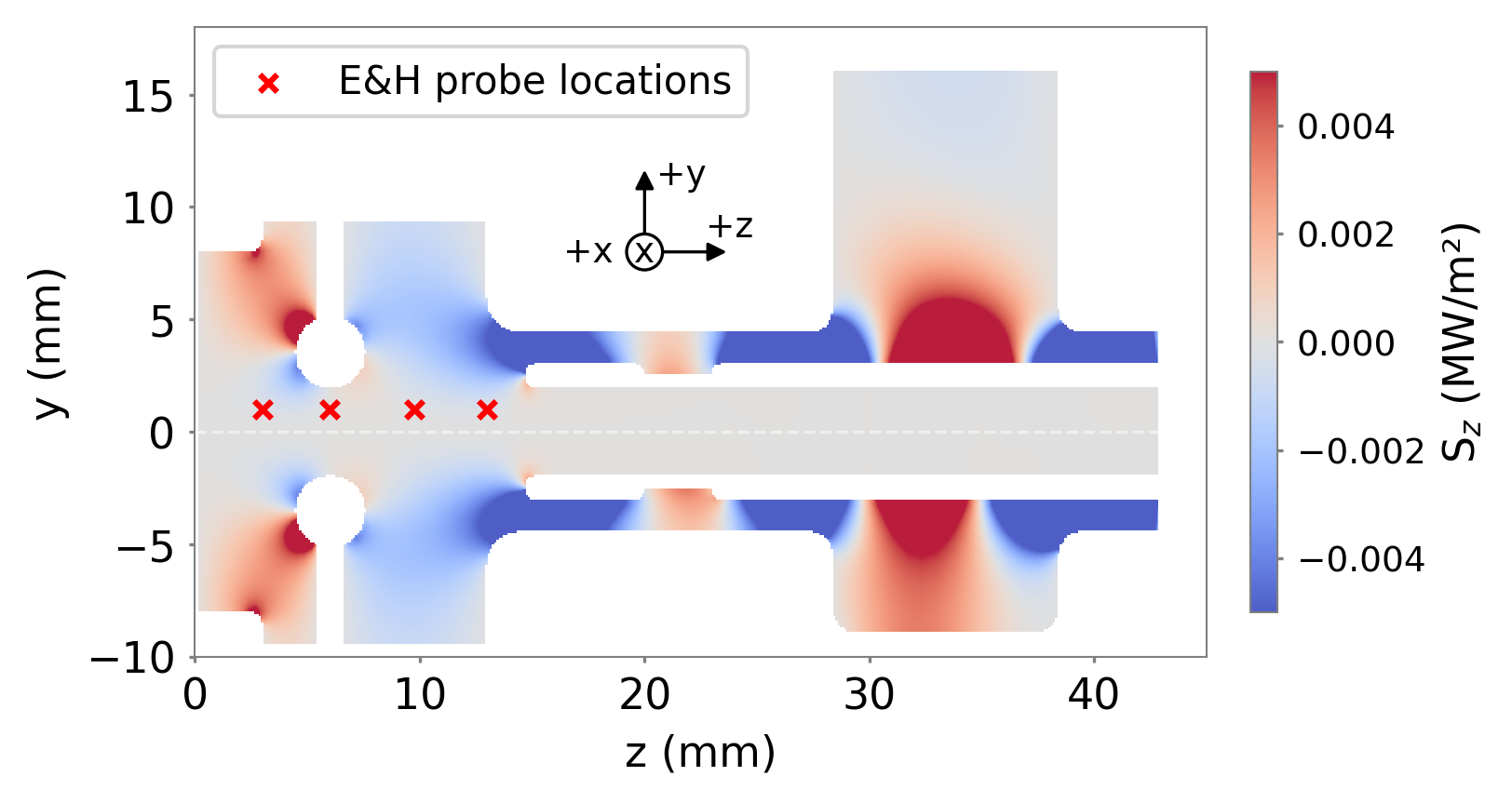}
    \caption{\justifying An example of Poynting vector distribution $S_z$ calculated at 2.8~ns in the $yz$ cross-section. Red cross marks indicate the E and H probe locations for computing Poynting vector as a function of time.}
    \label{fig:probe_locs}
\end{figure}

To capture this transient behavior, a detailed analysis of the EM fields was performed at different time slices throughout the entire RF pulse duration. The EM field was sampled at every 1° interval of the X-band frequency, which corresponds to a temporal resolution of $\sim 0.2$~ps. Additionally, for direct comparison, the steady-state profiles were scaled to match the peak field of the specific RF bucket and were phase-aligned to the corresponding time slices of the transient case, following the sinusoidal standing wave behavior described by Eq.~\ref{eq:standing_wave}.

Figure.~\ref{fig:Ez_in_diff_cycles}~(a) shows the field evolution in the 30$^\text{th}$ RF bucket, starting from 2.5684~ns during the initial filling stage, with field snapshots plotted at every 10°. The real-time $E_z$ simulated from the T-domain solver is shown as solid purple lines, while the steady-state field distribution from the F-domain solver at 11.68~GHz (shown in Fig.\ref{fig:Ez_F_domain}) is represented by red dashed lines. At this early stage, the field at most time slices deviates significantly from the steady-state behavior. For example, at phases such as 70°, 80°, 90°, and 100°, the on-axis field shows a ``0-mode”-like pattern, with the fields in both the half-cell and full-cell oscillating in phase. However, this behavior does not correspond to the actual 0-mode of the cavity, as the excitation bandwidth of the 9-ns trapezoidal RF pulse is $\sim 177$~MHz~\cite{chen2025design}, comparable to the ideal PETS signal shown in Fig.~\ref{fig:PETS_rf}, which is not wide enough to excite the 0-mode at 12.22~GHz (shown in Fig.~\ref{fig:S11}).

As the cavity gradually fills up, it reaches its peak field of 350~MV/m in the 88$^\text{th}$ RF bucket (starting at 7.489~ns). As shown in Fig.~\ref{fig:Ez_in_diff_cycles}~(b), the field at most time slices agrees well with the steady-state E$_z$, except around the ``polarity inversion phase" of 90°, where deviations are still observed in Fig.~\ref{fig:Ez_in_diff_cycles}~(c). For comparison, the field distributions from different RF buckets are presented, including the 80$^\text{th}$, 88$^\text{th}$ (peak field bucket), and 96$^\text{th}$ buckets during the transient state, and the 410$^\text{th}$ bucket (starting from 35.168~ns), which corresponds to the steady-state conditions, simulated using a 50-ns RF pulse. In Fig.~\ref{fig:Ez_in_diff_cycles}~(c), during the transient state, the fields in all RF buckets demonstrate similar behavior around the ``polarity inversion phase" at 90°, where the fields in the half-cell and full-cell oscillate in-phase, suggesting a ``0-mode"-like pattern. In contrast, the field in the $410^\text{th}$ bucket displays the expected SW behavior of the $\pi$-mode.

During transient states, given the deviations in $E_z$ from a typical SW cavity behavior, particularly the ``0-mode"-like pattern observed during the polarity inversion phase around 90° can likely be attributed to the dynamically changing RF power flow within the cavity. The Xgun, being a strongly over-coupled cavity with a coupling coefficient $\beta=22.3$ specifically designed to accommodate fast power filling for short-pulse operation, inherently results in significant power reflection. Additionally, the Xgun's design incorporates a one-sided coaxial feeding. Thus, during the transient state before the cavity is fully filled, RF power continuously redistributes between sections of the cavity, with both forwarded and reflected power interacting dynamically. Since the coupling coefficients between different cavity sections (i.e., half-cell to full-cell, and full-cell to coaxial coupler) are not equal, the RF power flowing into the cavity and dynamically reflecting back to the waveguide experiences different coupling across sections. As a result, different sections of the cavity respond differently to the RF filling process. This non-uniform power flow may lead to temporary field patterns that deviate from the designed $\pi$-mode. An extreme case could show a false ``0-mode" pattern, caused by faster field changes in one section of the cavity due to strong coupling, while other sections lag behind with a delayed response due to relatively weaker coupling. Once the cavity is fully filled and reaches steady-state conditions, the power flow and field distribution reach equilibrium, allowing the cavity to operate closer to a typical SW structure.

To quantify the RF power flow process and field build-up within the Xgun, the Poynting vector ($\vec{S}$), representing the directional energy flux density, was calculated at a few locations along the main power flow axis of the cavity. Field probes with locations summarized in Fig.~\ref{fig:probe_locs} to monitor $\vec{E}$ and $\vec{H}$ were added in the CST T-domain solver to compute the Poynting vector as $\vec{S}=\vec{E}\times \vec{H}$. Specifically, these field monitor probes were located in the middle of the half-cell at (1, 1, 3)~mm, the iris (1, 1, 6)~mm, the middle of the full-cell (1, 1, 9.75)~mm, and the full-cell exit (1, 1, 13) mm. A small offset (+1~mm) along both $x$ and $y$ axis was introduced since the Xgun geometry has a vanishing Poynting vector on its axis. The primary power flow within the cavity, which is along the $z$-axis, can be calculated as follows,
\begin{equation}
S_z(t)=E_x(t) H_y(t)-E_y(t) H_x(t).
\end{equation}

Figure~\ref{fig:Sz_total} illustrates the temporal evolution of the Poynting vector $S_z(t)$ at different locations, calculated for both the 9-ns and 50-ns RF pulses. Since $S_z(t)$ is evaluated from the $\vec{E}(t)$ and $\vec{H}(t)$ fields, which naturally oscillate in the RF photogun. To describe the power flow dynamics, the positive envelope ($S_z^+$) is computed as the rolling average of the waveform's positive peaks, representing the outward power flow over time. The negative envelope ($S_z^-$) is calculated in a similar way, based on the waveform's negative peaks, representing the inward power flow. The net power flow over time is then defined as the sum of the two components, $S_z =S_z^+ + S_z^-$.

For the 9-ns pulse, shown in Fig.~\ref{fig:Sz_total}~(a), the net $S_z$ at each location changes over time as expected from the transient-state behavior of the Xgun. The positive net $S_z$ indicates that power propagates along the $+z$ direction, corresponding to power reflecting out of the cavity, while the negative net $S_z$ indicates that power propagates in the $-z$ direction, representing power flowing into the cavity. At all the monitored locations, the iris and the full-cell exit are particularly notable: at the iris, the net $S_z$ indicates the power transfer between the half-cell and the full-cell, while at the full-cell exit, the net $S_z$ represents the interaction between the full-cell and the coaxial coupler (or effectively the power flow between the entire cavity and the coaxial coupler). At $\sim$7.5~ns, the net $S_z$ at all locations, particularly at the full-cell exit, crosses zero, indicating a reversed net power flow from inward to outward. At this moment, the cavity is mostly filled (though still in the transient state), and the field on cathode reaches its maximum, which is consistent with the results of the cathode field evolution shown in Fig.~\ref{fig:E_cathode}. Immediately after, the RF power within the cavity begins to decay. Additionally, before 7.5~ns, during the filling-up stage, the absolute value of the net $S_z$ at the full-cell exit is notably higher than at the iris, implying a stronger coupling between the full-cell and the coaxial coupler than between the full-cell and the half-cell. In other words, the full-cell responds to the RF variations more quickly, while the half-cell experiences a delayed response, resulting in a false ``0-mode"-like pattern around the polarity inversion phase at 90° or similarly 270°. 

For a ``long" 50-ns RF pulse, shown in Fig.~\ref{fig:Sz_total}~(b), after reaching steady-state conditions, starting from $\sim$23~ns, the net $S_z$ at all locations remains constant over time until it begins to decay. This constant net power flow is expected for a standing-wave (SW) cavity in a steady-state condition. In this state, the forward power, reflected power, and power dissipated into the cavity walls reach equilibrium and become well-balanced, providing a stable electromagnetic field distribution in the cavity.

Short-pulse operation of the Xgun allows the achievement of a high-gradient on the cathode surface, even during transient states when the cavity is not fully filled. Time-dependent field simulations demonstrate a wide operation range across tens of RF buckets, suitable for high-gradient beam tests. However, the non-equilibrium power flow leads to field distributions that deviate from typical SW cavity behavior, especially near polarity inversion phases (90° or 270°), though the field magnitude is low during these phases. The next section explores the influence of these transient RF characteristics on beam dynamics.

\section{Beam Dynamics in the Transient States}

The transient behavior of the RF fields, as analyzed in the previous section, can directly impact the beam dynamics. To quantify this impact, this section focuses on preliminary beam dynamics simulations of the Xgun to gain insight into the transient effects on basic beam properties: beam kinetic energy, transverse emittance, and rms bunch length. To accurately account for the RF cavity's transient behavior, a new simulation approach is adopted. Since traditional beam simulations implementing standing-wave (SW) cavities typically require a single 1D or 3D field map, where the field is scaled to a user-defined amplitude and follows a temporal oscillation described as $E_{\text{0}} \cdot \cos(2\pi f t)$ (or as $sin$ convention). While this method effectively models SW cavities, it cannot capture the transient RF behavior of the Xgun. To address this, a dynamic field mapping approach is implemented, which continuously updates the time-dependent field maps during the beam-cavity interaction, providing an accurate real-time cavity field distribution.

\subsection{Simulation Setup}

To investigate the evolution of beam dynamics, we utilize the particle-tracking program \verb|ASTRA|. The program represents the electron-beam distribution as a pre-defined macro-particle ensemble and integrates the equation of motion in user-defined electromagnetic fields described. The field can be represented either as an axial 1D field profile $E_{z,0}(z)$, or as a full 3D Cartesian map for each of the 6 components $(\vec E, \vec B)$ associated with the map. As each macro-particle passes through the cavity, its position and time of arrival will determine the field it experiences. In \verb|ASTRA|, simplifying to the case of a 1D field map, the cavity's electric field experienced by a macro-particle is 
\begin{equation} \label{eq:astra1dsimple}
    E_z(z,t)=E_{\text{z,0}}(z) \cdot \text{sin}(2\pi f t_i + \phi),
\end{equation}
where $E_{\text{z,0}}(z)$ is the field amplitude, $f$ is the cavity frequency,  $\phi$ is the given launching phase, and $t_i$ is a particle-specific local time of arrival. Additionally, \verb|ASTRA| follows the $sin$ convention for defining the electric field phase variation, which is different from CST. Thus, in an ideal SW cavity, the oscillation of the RF field remains strictly periodic, governed by the specified frequency and initial launch phase. If the macro-particle's arrival time is known, the corresponding field it experiences can be directly determined. 

For the Xgun operating in transient states, the RF field is not strictly periodic so that the harmonic time dependence assumed in Eq.~\ref{eq:astra1dsimple} does not apply. Instead, the field experienced by a particle is dynamically changing over time. To accurately model this transient behavior, a new simulation workflow is implemented using \verb|ASTRA|, where the RF field is synchronized with the particle's arrival time and continuously updated during the tracking process. 
\begin{figure}[t]
\centering
\includegraphics[width=0.95\linewidth]{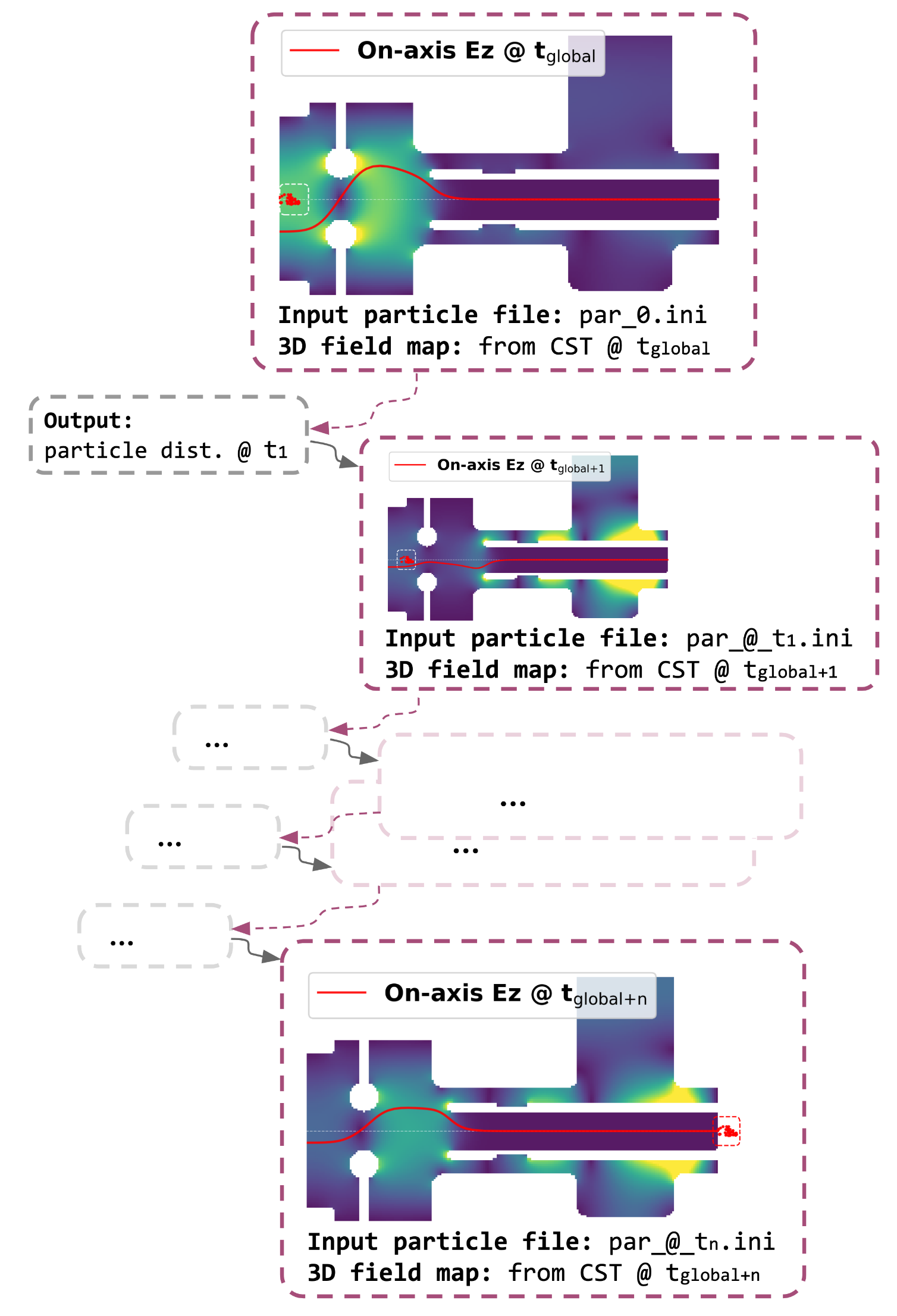}
\caption{\justifying Simulation workflow for particle tracking in transient RF fields.}
\label{fig:beam_sim_workflow}
\end{figure}
Figure~\ref{fig:beam_sim_workflow} illustrates the simulation workflow for tracking particles within transient RF fields. The transient RF field maps are exported from the CST T-domain, sampled at 1° intervals in phase, corresponding to a time step of $\sim$0.2~ps at 11.68~GHz. This resolution represents the smallest time step practically exportable from the CST T-domain solver while maintaining sufficient accuracy in capturing rapid RF field variations. For a beam launched in the $n^\text{th}$ RF cycle at a phase of $\phi _0$, the corresponding global time is calculated as
\begin{equation}
    t_{\text{global}} = n \cdot T_{RF} +  \frac{\phi_0}{360} \cdot T_{RF},
\end{equation}
where $T_{RF}$ is the RF period. Once the field map is located, particle tracking in \verb|ASTRA| starts with $t=0$, launched with the field map at $t_{\text{global}}$. 
As the simulation progresses, \verb|ASTRA| performs particle tracking with a time step of approximately 0.2 ps, continuously updating the external field map to correspond with the synchronized time. This iterative process continues until the reference particle reaches the cavity exit at $z=50$~mm. To provide a reference and quantify the computational workload, we estimate the number of field updates required for the one set of simulations. Assuming the particle travels at the speed of light, its time of flight from the cathode surface to the 50-mm exit is of $\sim 0.17$~ns, spanning slightly less than 2 RF cycles at 11.68 GHz. Given the field sampling resolution of 1°, this results in over 700 field map updates during the tracking process.

\subsection{Beam Properties in Transient RF Fields}
To gain insights into the impact of transient fields on the beam dynamics, we examine the evolution of the beam's kinetic energy ($E_{kin}$), transverse emittance ($\varepsilon$) and rms bunch length ($\sigma{_z}$) as a function of the launching phases $\phi \in [0°, 360°]$ with a scanning step size of 5°. All the parameters are recorded downstream of the Xgun cavity (corresponding to $z\simeq 50$~mm). To isolate the impact of transient fields and eliminate other potential contributions in the beam dynamics simulations, the following assumptions were made: ($i$) the initial thermal emittance was set to zero, ($ii$) space-charge including image-charge effects at the photocathode were neglected, and ($iii$) no external solenoidal-lens focusing was implemented. The beam-dynamics simulations employed 3D Cartesian field maps to model the Xgun recorded over a computational domain of $\delta x\times \delta y \times \delta z = 6\times 6 \times 50$~mm$^3$ transversely centered on the Xgun axis. The map's spatial extent ensures the macro-particle remains within the map while minimizing the data size for computational efficiency, The number of grid points $N_x\times N_y \times N_z = 21\times 21\times 144$ was selected to provide sufficient spatial resolution for beam tracing. For comparison purposes, beam simulations were conducted over multiple RF cycles, corresponding to the plotted field shown in Fig.\ref{fig:Ez_in_diff_cycles}: $\#$80, $\#$88 (the RF cycle where the field amplitude on the photocathode reaches its maximum value), and $\#$96, as well as a steady-state field map produced from CST's F-solver to simulate the case of an ideal SW cavity in \verb|ASTRA|. Additionally, in transient states, as the RF field continuously builds up, the peak field amplitudes change from cycle to cycle. To focus exclusively on the effects of transient field patterns (independent of amplitude variations), the peak field in each RF cycle was normalized to a consistent 350~MV/m. 

\begin{figure}[t]
    \centering
    \includegraphics[width=0.95\linewidth]{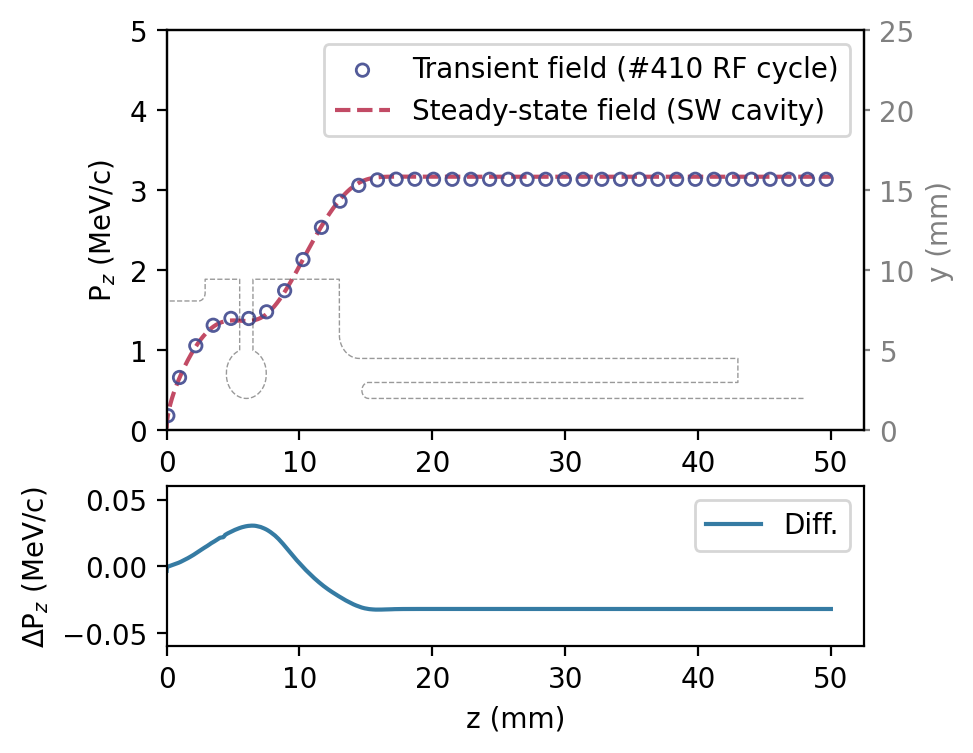}
    \caption{\justifying Top: longitudinal momentum ($P_z$) as a function of $z$ at a fixed launching phase of 270°. Dashed line: tracking with the standing-wave cavity field. Scattered points: tracking with the transient field maps. Thin dashed line (light gray): Xgun geometry. Bottom: residual ($\Delta P_z$) between the two data curves on the top.}
    \label{fig:beam_sim_pz}
\end{figure}

As a first step in confirming the validity of the dynamic field mapping workflow described in the previous section, the beam's longitudinal momentum ($P_z$) was studied as a function of $z$ at a fixed launching phase of 270°, capturing the complete tracking process in the Xgun. For comparison, the \#410 RF cycle (steady-state regime) was compared with a simulation employing a ideal SW cavity field in \verb|ASTRA|. As shown in Fig.~\ref{fig:beam_sim_pz} (top), the $P_z$ evolution in the Xgun cavity, tracked using transient field maps from the $\#$410 RF cycle (under steady-state conditions) using the dynamic field mapping workflow, demonstrates good agreement with tracking results using the ideal SW cavity model. The maximum fractional discrepancy is $\le 1$\%; see Fig.~\ref{fig:beam_sim_pz} (lower plot). 

\begin{figure}[t]
    \centering
    \includegraphics[width=0.95\linewidth]{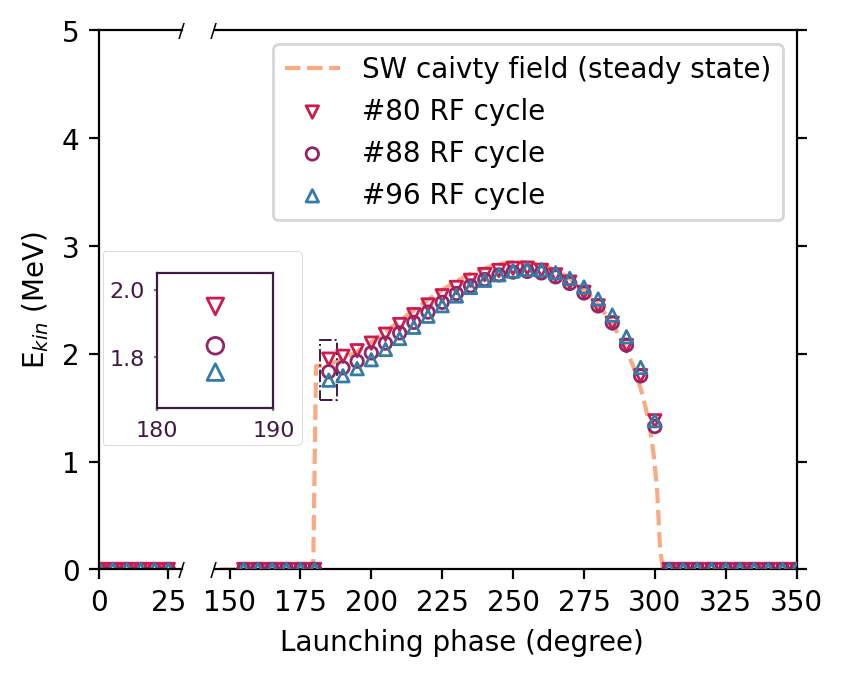}
    \caption{\justifying Kinetic energy ($E_{kin}$) as a function of launching phase at the Xgun exit. Zoomed-in inset shows $E_{kin}$ at 185°, where the largest discrepancy between the transient and steady-state results is observed.}
    \label{fig:beam_sim_Ekin}
\end{figure}

\begin{figure}[t]
    \centering
    \begin{subfigure}[b]{\linewidth}
        \includegraphics[width=0.95\linewidth]{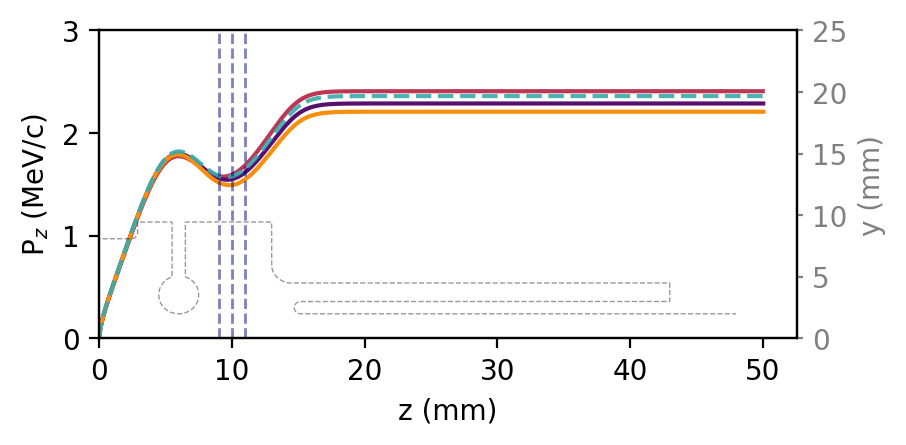}
        \caption{Evolution of the longitudinal momentum $P_z$ along the beamline distance $z$.}
    \end{subfigure}
    \hfill
    \begin{subfigure}[b]{\linewidth}
        \includegraphics[width=0.9\linewidth]{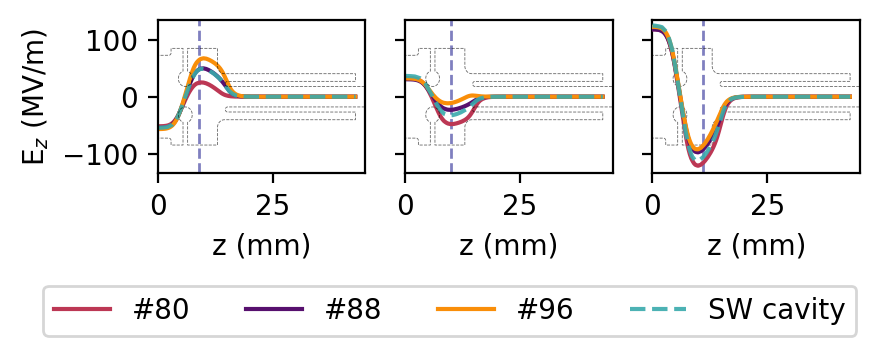}
        \caption{On-axis $E_z$ at different location slices.}
    \end{subfigure}
    \caption{\justifying At a launching phase of 185°: (a) $P_z$ as a function of $z$, comparing the results from dynamic field mapping workflow (for RF cycles $\#80$, $\#88$, and $\#96$) with the result from the SW cavity field. (b) Snapshots of the on-axis $E_z$ field at the time when the reference particle crosses the longitudinal location corresponding to each of the vertical dashed lines appearing in (a).}
    \label{fig:beam_sim_E}
\end{figure}

Following the established simulation workflow, Fig.~\ref{fig:beam_sim_Ekin} presents the beam's kinetic energy at the Xgun exit as a function of the launching phase, scanned across three separate RF cycles. The $E_{kin}$ evolution in the transient states, simulated using the dynamic field mapping method, generally agrees well with the results obtained using the pure SW cavity model, showing only minor deviations. The most significant discrepancy between the dynamic and steady-state SW models occurs at a launching phase of 185°. 
To understand this discrepancy, the longitudinal momentum ($P_z$) of the beam as a function of position ($z$) at the 185° launching phase was studied in detail. As shown in Fig.~\ref{fig:beam_sim_E}~(a), notably, $P_z$ begins to diverge once the particle propagates approximately 9~mm into the cavity. To further investigate how the local electric field ($E_z$) affects this divergence, we examine the on-axis $E_z$ experienced by the reference particle at it reaches three different longitudinal positions (9~mm, 10~mm, and 11~mm), marked by the vertical dashed lines in Fig.~\ref{fig:beam_sim_E}~(a). Figure~\ref{fig:beam_sim_E}~(b) presents the electric field experienced by the reference particle at these positions. The results reveal a clear trend: at these positions, RF cycle $\#$80 exhibits the highest transient accelerating gradient (or the least decelerating gradient) compared to other RF cycles, correlating with the higher energy gain observed in Fig.~\ref{fig:beam_sim_Ekin}. 

\begin{figure}[t]
    \center
    \includegraphics[width=0.95\linewidth]{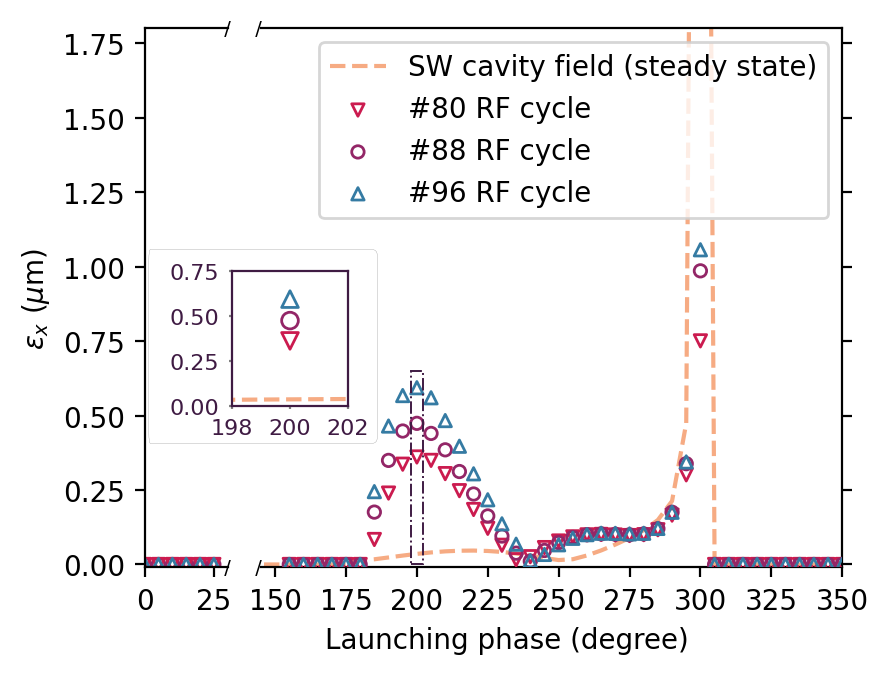}
    \caption{\justifying Transverse emittance x ($\varepsilon{_x}$) as a function of launching phase at the Xgun exit. Zoomed-in inset shows $\varepsilon{_x}$ at 200°, where the largest discrepancy is observed.}
    \label{fig:beam_sim_emit}
\end{figure}

Similarly, the transverse emittance ($\varepsilon_x$) evolution was also studied in the transient RF fields as shown in Fig.~\ref{fig:beam_sim_emit}. Here, the initial RMS beam size was set to 0.1~mm, and the bunch length was 300~fs (FWHM).

\begin{figure}[t]
    \centering
    \includegraphics[width=0.95\linewidth]{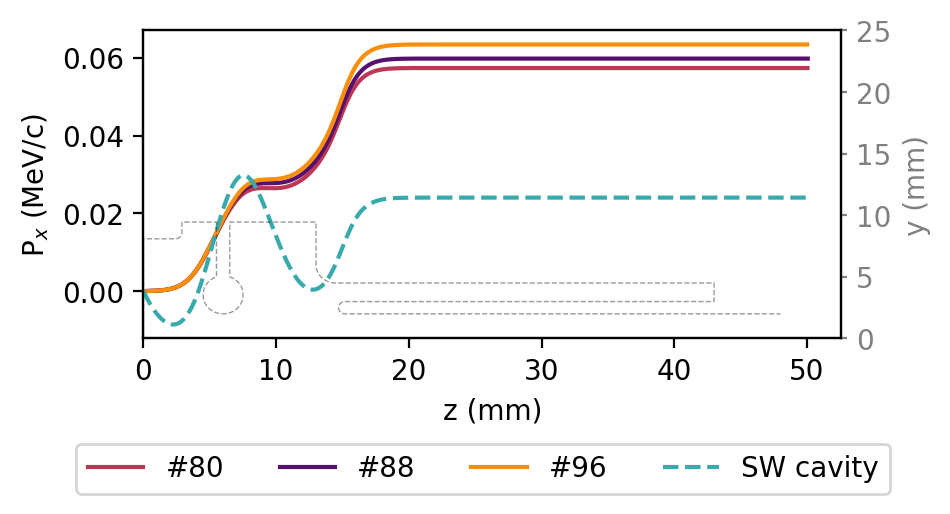}
    \caption{\justifying At a launching phase of 200°, $P_x$ as a function of $z$, comparing the results from dynamic field mapping workflow (for RF cycles $\#80$, $\#88$, and $\#96$) with the result from the SW cavity field.}
    \label{fig:beam_sim_Px}
\end{figure}

As shown in Fig.~\ref{fig:beam_sim_emit}, the evolution of $\varepsilon_x$ as a function of the launching phase demonstrates that transient field patterns significantly contribute to emittance behavior, which is different from the results simulated by the SW model. While the global minimum emittance ($\sim$10~nm) in transient fields aligns with the results from the SW model, the optimal launching phase shifts from the $\sim$250° (SW cavity) to $\sim$240° (transient fields). This indicates that low emittance can still be achieved in transient states even when the cavity stays underfilled. Furthermore, at a launching phase of 200°, where the largest discrepancy in $\varepsilon_x$ is observed, Fig.~\ref{fig:beam_sim_Px} shows that for the single reference particle, $P_x$ is notably higher in the transient‐field simulations compared to the SW model. This higher transverse momentum indicates that transient fields can introduce large transverse kicks at certain phases, which may directly contribute to emittance growth. Overall, these observations highlight the importance of considering transient fields in future beam optimizations, as the optimal launching phase may depend on the cavity's transient behavior.

A comprehensive analysis of the beam dynamics would require accounting for multiple effects including space charge effects, and implementing an external solenoidal field for emittance compensation~\cite{CARLSTEN1989313}. In this paper we deliberately exclude these factors and focus solely on gaining insights into the impact of transient field patterns on the beam dynamics (and particularly on the transverse emittance). 

\begin{figure}[h]
    \center
    \includegraphics[width=0.95\linewidth]{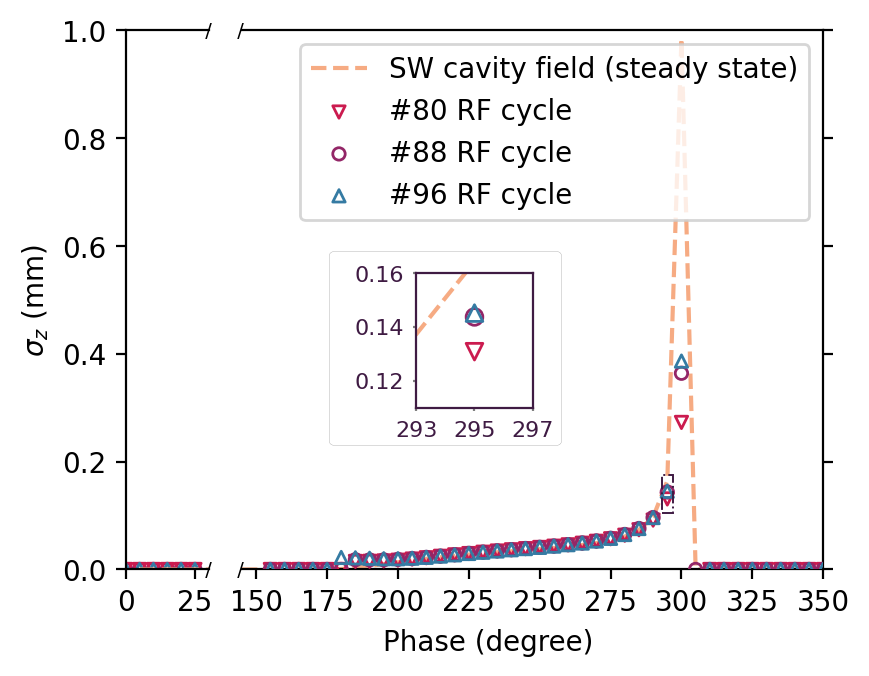}
    \caption{\justifying Bunch length ($\sigma{_z}$) as a function of launching phase at the Xgun exit. Zoomed-in inset shows $\sigma{_z}$ at 290°.}
    \label{fig:beam_sim_bunch_len}
\end{figure}

Figure~\ref{fig:beam_sim_bunch_len} gives the rms bunch length ($\sigma{_z}$) evolution at different launching phases. The bunch length shows minimal sensitivity to transient fields, and mostly agrees with the results simulated using the SW field, with minor discrepancies at a low-energy phase of $\sim$295°.

For future work, a potential improvement would be increasing the resolution of the dynamic field updates, as the current limitation of updating the field at 1° (~0.2~ps) per iteration may still result in non-smooth transitions in the RF field. Such abrupt changes in the field could introduce discontinuities in the beam motion when simulated in \verb|ASTRA|, potentially affecting the accuracy of the results. However, based on the current computational resources, updating the field at 1° ($\sim$0.2~ps) increments represents the highest resolution we can achieve, which is limited by the capabilities of the CST software. Additionally, we have also considered the possibility of fitting the approximately 700 RF field map datasets required for a single simulation to a superposition of multiple standing wave cavities, expressed as
\begin{equation}
    E(z, t)=\sum_n E_{z, n}(z) \sin \left(2 \pi f_n t\right).
\end{equation}
This approach would allow the transient fields to be modeled in a manner similar to a standard SW cavity in \verb|ASTRA| simulations. However, due to the non-monotonic behavior observed in the system, the fitting process is not straightforward and may not fully capture the complexities of the transient RF fields. Nevertheless, this presents a potential way for future work to improve the efficiency of the modeling of transient effects in beam simulations.

\section{Conclusion}

In this work, we have conducted a detailed study of the transient RF characteristics of the Xgun with a particular focus on the field build-up process and its influence on beam dynamics. Our analysis of RF power flow within the cavity revealed a non-uniform power distribution during the transient state, leading to field patterns that deviate from the ideal SW cavity behavior. Beam dynamics simulations demonstrated that the transient RF fields have an insignificant impact on beam energy and bunch length when compared to the SW cavity. Though the emittance trend shows a large discrepancy between the transient RF fields and the SW cavity field at various launching phases, the global minimum emittance remains comparable to that achieved with the SW cavity field. The developed tool will be further utilized to optimize beam dynamics in an integrated photoinjector, while also enabling the exploration of the interplay between the drive-beam shape pattern and the beam dynamics of the accelerated bunch. This capability also opens new avenues for controlling and optimizing the field pattern and beam dynamics in accelerating structures operating in the short-pulse regime relevant for future light source or energy-frontier colliders based on the TBA scheme.

\begin{acknowledgments}
This research was supported by Laboratory Directed Research and Development (LDRD) funding from Argonne National Laboratory, provided by the Director, Office of Science, U.S. Department of Energy, under Contract No. DE-AC02-06CH11357. 
\end{acknowledgments}

\bigskip
\appendix

\section{AWA Ultraviolet Laser-Pulse Splitter}\label{beam_splitter}

The drive-beam bunch train required for the Two-Beam Acceleration (TBA) technique is generated via photoemission in an L-band RF photoinjector. To produce multiple bunches at the L-band AWA gun frequency ($f_L=1.3$~GHz), the ultraviolet laser pulse that triggers photoemission is structured as a train of eight successive pulses, each separated by $T_b=1/(1.3\times 10^9)=769$~ps. 

\begin{widetext}
\begin{figure*}[h!]
    \centering
    \includegraphics[width=0.7\linewidth]{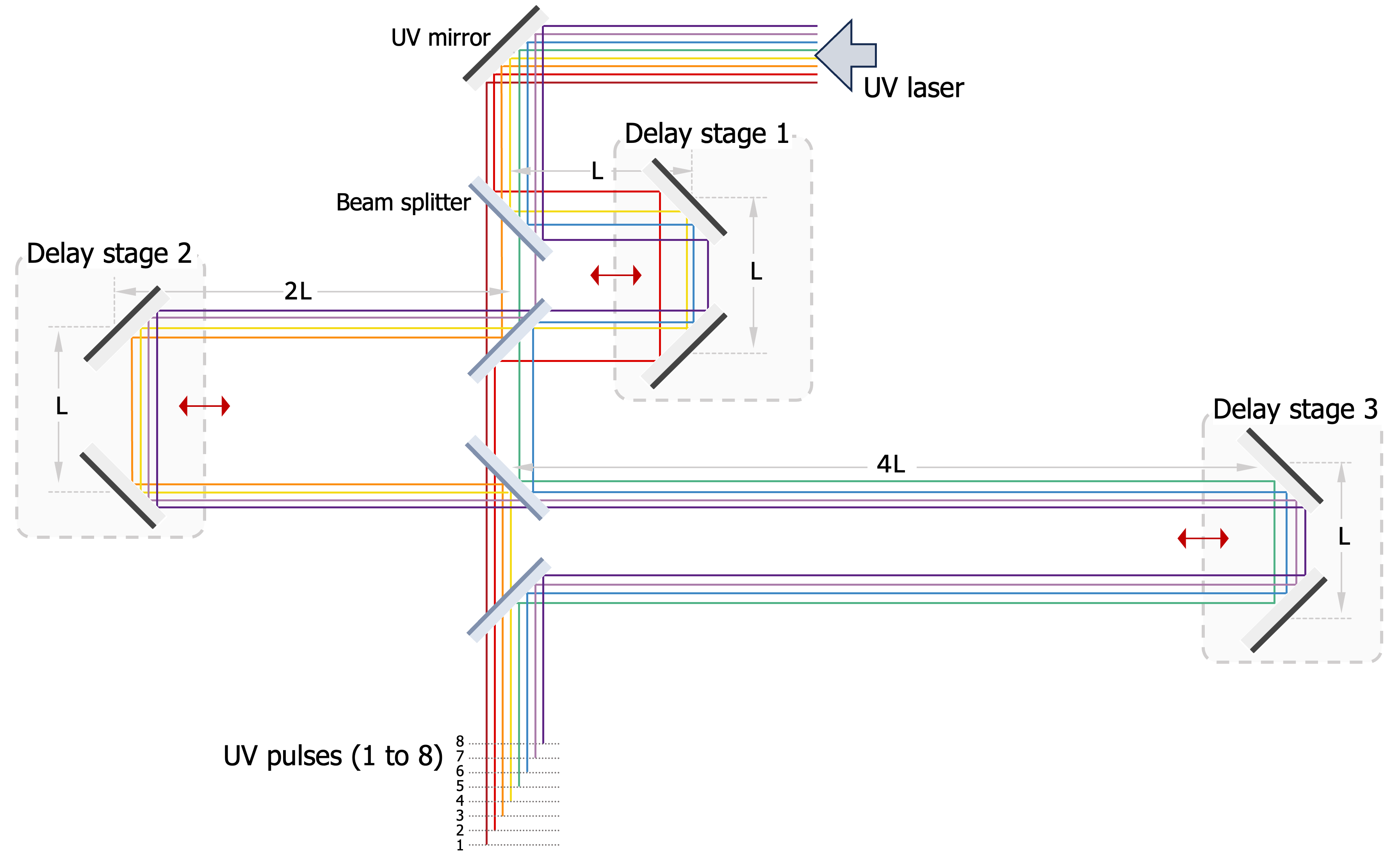}
    \caption{Schematic diagram of AWA's laser-pulse splitter system for the generation of multiple ultraviolet pulses. The distance $L\simeq 11.53$~cm, and each beam splitter has an equal transmittance and reflectance of 50\%. The colored lines represent the same optical ray and are displaced for clarity.}
    \label{fig:delay_stage}
\end{figure*}
\end{widetext}

Figure~\ref{fig:delay_stage} shows the generation of multiple UV pulses with controlled delays using optical components. For the final pulse train, adjacent pulses are separated by $2L=c/f_L\simeq 23.06$~cm, corresponding to the temporal separation of 769~ps. In this setup, pulse 1 serves as the reference, with the rest bunches aligned to it. Pulse 2, 3, and 5 are individually adjustable via dedicated delay stages 1, 2, and 3. The remaining pulses (4, 6, 7, and 8) inherit fixed delays from the preconfigured optical system. For example, a 1° phase shift ($\sim$2~ps at 1.3~GHz) added on pulse 2 (controlled by delay stage 1) introduces a final phase shift across the train: (0°, 1°, 0°, 1°, 0°, 1°, 0°, 1°). 

\clearpage

\begin{thebibliography}{20}%
\makeatletter
\providecommand \@ifxundefined [1]{%
 \@ifx{#1\undefined}
}%
\providecommand \@ifnum [1]{%
 \ifnum #1\expandafter \@firstoftwo
 \else \expandafter \@secondoftwo
 \fi
}%
\providecommand \@ifx [1]{%
 \ifx #1\expandafter \@firstoftwo
 \else \expandafter \@secondoftwo
 \fi
}%
\providecommand \natexlab [1]{#1}%
\providecommand \enquote  [1]{``#1''}%
\providecommand \bibnamefont  [1]{#1}%
\providecommand \bibfnamefont [1]{#1}%
\providecommand \citenamefont [1]{#1}%
\providecommand \href@noop [0]{\@secondoftwo}%
\providecommand \href [0]{\begingroup \@sanitize@url \@href}%
\providecommand \@href[1]{\@@startlink{#1}\@@href}%
\providecommand \@@href[1]{\endgroup#1\@@endlink}%
\providecommand \@sanitize@url [0]{\catcode `\\12\catcode `\$12\catcode `\&12\catcode `\#12\catcode `\^12\catcode `\_12\catcode `\%12\relax}%
\providecommand \@@startlink[1]{}%
\providecommand \@@endlink[0]{}%
\providecommand \url  [0]{\begingroup\@sanitize@url \@url }%
\providecommand \@url [1]{\endgroup\@href {#1}{\urlprefix }}%
\providecommand \urlprefix  [0]{URL }%
\providecommand \Eprint [0]{\href }%
\providecommand \doibase [0]{https://doi.org/}%
\providecommand \selectlanguage [0]{\@gobble}%
\providecommand \bibinfo  [0]{\@secondoftwo}%
\providecommand \bibfield  [0]{\@secondoftwo}%
\providecommand \translation [1]{[#1]}%
\providecommand \BibitemOpen [0]{}%
\providecommand \bibitemStop [0]{}%
\providecommand \bibitemNoStop [0]{.\EOS\space}%
\providecommand \EOS [0]{\spacefactor3000\relax}%
\providecommand \BibitemShut  [1]{\csname bibitem#1\endcsname}%
\let\auto@bib@innerbib\@empty
\bibitem [{\citenamefont {Emma}\ \emph {et~al.}(2010)\citenamefont {Emma}, \citenamefont {Akre}, \citenamefont {Arthur}, \citenamefont {Bionta}, \citenamefont {Bostedt}, \citenamefont {Bozek}, \citenamefont {Brachmann}, \citenamefont {Bucksbaum}, \citenamefont {Coffee}, \citenamefont {Decker} \emph {et~al.}}]{emma2010first}%
  \BibitemOpen
  \bibfield  {author} {\bibinfo {author} {\bibfnamefont {P.}~\bibnamefont {Emma}}, \bibinfo {author} {\bibfnamefont {R.}~\bibnamefont {Akre}}, \bibinfo {author} {\bibfnamefont {J.}~\bibnamefont {Arthur}}, \bibinfo {author} {\bibfnamefont {R.}~\bibnamefont {Bionta}}, \bibinfo {author} {\bibfnamefont {C.}~\bibnamefont {Bostedt}}, \bibinfo {author} {\bibfnamefont {J.}~\bibnamefont {Bozek}}, \bibinfo {author} {\bibfnamefont {A.}~\bibnamefont {Brachmann}}, \bibinfo {author} {\bibfnamefont {P.}~\bibnamefont {Bucksbaum}}, \bibinfo {author} {\bibfnamefont {R.}~\bibnamefont {Coffee}}, \bibinfo {author} {\bibfnamefont {F.-J.}\ \bibnamefont {Decker}}, \emph {et~al.},\ }\bibfield  {title} {\bibinfo {title} {First lasing and operation of an {\aa}ngstrom-wavelength free-electron laser},\ }\href@noop {} {\bibfield  {journal} {\bibinfo  {journal} {nature photonics}\ }\textbf {\bibinfo {volume} {4}},\ \bibinfo {pages} {641} (\bibinfo {year} {2010})}\BibitemShut {NoStop}%
\bibitem [{\citenamefont {Piot}\ \emph {et~al.}(2024)\citenamefont {Piot}, \citenamefont {Chen}, \citenamefont {Frame}, \citenamefont {Jing}, \citenamefont {Kuzikov}, \citenamefont {Lu},\ and\ \citenamefont {Power}}]{piot:fls2023-mo4c2}%
  \BibitemOpen
  \bibfield  {author} {\bibinfo {author} {\bibfnamefont {P.}~\bibnamefont {Piot}}, \bibinfo {author} {\bibfnamefont {G.}~\bibnamefont {Chen}}, \bibinfo {author} {\bibfnamefont {E.}~\bibnamefont {Frame}}, \bibinfo {author} {\bibfnamefont {C.-J.}\ \bibnamefont {Jing}}, \bibinfo {author} {\bibfnamefont {S.}~\bibnamefont {Kuzikov}}, \bibinfo {author} {\bibfnamefont {X.}~\bibnamefont {Lu}},\ and\ \bibinfo {author} {\bibfnamefont {J.}~\bibnamefont {Power}},\ }\bibfield  {title} {\bibinfo {title} {Development of a compact light source using a two-beam-acceleration technique},\ }in\ \href {https://doi.org/10.18429/JACoW-FLS2023-MO4C2} {\emph {\bibinfo {booktitle} {Proc. 67th ICFA Adv. Beam Dyn. Workshop Future Light Sources (FLS'23)}}},\ \bibinfo {series and number} {\bibinfo {series} {ICFA Advanced Beam Dynamics Workshop}\ No.~\bibinfo {number} {67}}\ (\bibinfo  {publisher} {JACoW Publishing, Geneva, Switzerland},\ \bibinfo {year} {2024})\ pp.\ \bibinfo {pages} {42--45}\BibitemShut {NoStop}%
\bibitem [{\citenamefont {Graves}\ \emph {et~al.}(2014)\citenamefont {Graves}, \citenamefont {Bessuille}, \citenamefont {Brown}, \citenamefont {Carbajo}, \citenamefont {Dolgashev}, \citenamefont {Hong}, \citenamefont {Ihloff}, \citenamefont {Khaykovich}, \citenamefont {Lin}, \citenamefont {Murari} \emph {et~al.}}]{graves2014compact}%
  \BibitemOpen
  \bibfield  {author} {\bibinfo {author} {\bibfnamefont {W.}~\bibnamefont {Graves}}, \bibinfo {author} {\bibfnamefont {J.}~\bibnamefont {Bessuille}}, \bibinfo {author} {\bibfnamefont {P.}~\bibnamefont {Brown}}, \bibinfo {author} {\bibfnamefont {S.}~\bibnamefont {Carbajo}}, \bibinfo {author} {\bibfnamefont {V.}~\bibnamefont {Dolgashev}}, \bibinfo {author} {\bibfnamefont {K.-H.}\ \bibnamefont {Hong}}, \bibinfo {author} {\bibfnamefont {E.}~\bibnamefont {Ihloff}}, \bibinfo {author} {\bibfnamefont {B.}~\bibnamefont {Khaykovich}}, \bibinfo {author} {\bibfnamefont {H.}~\bibnamefont {Lin}}, \bibinfo {author} {\bibfnamefont {K.}~\bibnamefont {Murari}}, \emph {et~al.},\ }\bibfield  {title} {\bibinfo {title} {Compact x-ray source based on burst-mode inverse compton scattering at 100 khz},\ }\href@noop {} {\bibfield  {journal} {\bibinfo  {journal} {Physical Review Special Topics-Accelerators and Beams}\ }\textbf {\bibinfo {volume} {17}},\ \bibinfo {pages} {120701} (\bibinfo {year} {2014})}\BibitemShut {NoStop}%
\bibitem [{\citenamefont {Qi}\ \emph {et~al.}(2020)\citenamefont {Qi}, \citenamefont {Ma}, \citenamefont {Zhao}, \citenamefont {Cheng}, \citenamefont {Jiang}, \citenamefont {Lu}, \citenamefont {Jiang}, \citenamefont {Qian}, \citenamefont {Wang}, \citenamefont {Zhang} \emph {et~al.}}]{qi2020breaking}%
  \BibitemOpen
  \bibfield  {author} {\bibinfo {author} {\bibfnamefont {F.}~\bibnamefont {Qi}}, \bibinfo {author} {\bibfnamefont {Z.}~\bibnamefont {Ma}}, \bibinfo {author} {\bibfnamefont {L.}~\bibnamefont {Zhao}}, \bibinfo {author} {\bibfnamefont {Y.}~\bibnamefont {Cheng}}, \bibinfo {author} {\bibfnamefont {W.}~\bibnamefont {Jiang}}, \bibinfo {author} {\bibfnamefont {C.}~\bibnamefont {Lu}}, \bibinfo {author} {\bibfnamefont {T.}~\bibnamefont {Jiang}}, \bibinfo {author} {\bibfnamefont {D.}~\bibnamefont {Qian}}, \bibinfo {author} {\bibfnamefont {Z.}~\bibnamefont {Wang}}, \bibinfo {author} {\bibfnamefont {W.}~\bibnamefont {Zhang}}, \emph {et~al.},\ }\bibfield  {title} {\bibinfo {title} {Breaking 50 femtosecond resolution barrier in mev ultrafast electron diffraction with a double bend achromat compressor},\ }\href@noop {} {\bibfield  {journal} {\bibinfo  {journal} {Physical review letters}\ }\textbf {\bibinfo {volume} {124}},\ \bibinfo {pages} {134803} (\bibinfo {year} {2020})}\BibitemShut {NoStop}%
\bibitem [{\citenamefont {Li}\ and\ \citenamefont {Musumeci}(2014)}]{li2014single}%
  \BibitemOpen
  \bibfield  {author} {\bibinfo {author} {\bibfnamefont {R.}~\bibnamefont {Li}}\ and\ \bibinfo {author} {\bibfnamefont {P.}~\bibnamefont {Musumeci}},\ }\bibfield  {title} {\bibinfo {title} {Single-shot mev transmission electron microscopy with picosecond temporal resolution},\ }\href@noop {} {\bibfield  {journal} {\bibinfo  {journal} {Physical Review Applied}\ }\textbf {\bibinfo {volume} {2}},\ \bibinfo {pages} {024003} (\bibinfo {year} {2014})}\BibitemShut {NoStop}%
\bibitem [{\citenamefont {Filippetto}\ \emph {et~al.}(2014)\citenamefont {Filippetto}, \citenamefont {Musumeci}, \citenamefont {Zolotorev},\ and\ \citenamefont {Stupakov}}]{filippetto2014maximum}%
  \BibitemOpen
  \bibfield  {author} {\bibinfo {author} {\bibfnamefont {D.}~\bibnamefont {Filippetto}}, \bibinfo {author} {\bibfnamefont {P.}~\bibnamefont {Musumeci}}, \bibinfo {author} {\bibfnamefont {M.}~\bibnamefont {Zolotorev}},\ and\ \bibinfo {author} {\bibfnamefont {G.}~\bibnamefont {Stupakov}},\ }\bibfield  {title} {\bibinfo {title} {Maximum current density and beam brightness achievable by laser-driven electron sources},\ }\href@noop {} {\bibfield  {journal} {\bibinfo  {journal} {Physical Review Special Topics-Accelerators and Beams}\ }\textbf {\bibinfo {volume} {17}},\ \bibinfo {pages} {024201} (\bibinfo {year} {2014})}\BibitemShut {NoStop}%
\bibitem [{\citenamefont {Grudiev}\ \emph {et~al.}(2009)\citenamefont {Grudiev}, \citenamefont {Calatroni},\ and\ \citenamefont {Wuensch}}]{grudiev2009new}%
  \BibitemOpen
  \bibfield  {author} {\bibinfo {author} {\bibfnamefont {A.}~\bibnamefont {Grudiev}}, \bibinfo {author} {\bibfnamefont {S.}~\bibnamefont {Calatroni}},\ and\ \bibinfo {author} {\bibfnamefont {W.}~\bibnamefont {Wuensch}},\ }\bibfield  {title} {\bibinfo {title} {New local field quantity describing the high gradient limit of accelerating structures},\ }\href@noop {} {\bibfield  {journal} {\bibinfo  {journal} {Physical Review Special Topics—Accelerators and Beams}\ }\textbf {\bibinfo {volume} {12}},\ \bibinfo {pages} {102001} (\bibinfo {year} {2009})}\BibitemShut {NoStop}%
\bibitem [{\citenamefont {Tan}\ \emph {et~al.}(2022)\citenamefont {Tan}, \citenamefont {Antipov}, \citenamefont {Doran}, \citenamefont {Ha}, \citenamefont {Jing}, \citenamefont {Knight}, \citenamefont {Kuzikov}, \citenamefont {Liu}, \citenamefont {Lu}, \citenamefont {Piot} \emph {et~al.}}]{tan2022demonstration}%
  \BibitemOpen
  \bibfield  {author} {\bibinfo {author} {\bibfnamefont {W.}~\bibnamefont {Tan}}, \bibinfo {author} {\bibfnamefont {S.}~\bibnamefont {Antipov}}, \bibinfo {author} {\bibfnamefont {D.}~\bibnamefont {Doran}}, \bibinfo {author} {\bibfnamefont {G.}~\bibnamefont {Ha}}, \bibinfo {author} {\bibfnamefont {C.}~\bibnamefont {Jing}}, \bibinfo {author} {\bibfnamefont {E.}~\bibnamefont {Knight}}, \bibinfo {author} {\bibfnamefont {S.}~\bibnamefont {Kuzikov}}, \bibinfo {author} {\bibfnamefont {W.}~\bibnamefont {Liu}}, \bibinfo {author} {\bibfnamefont {X.}~\bibnamefont {Lu}}, \bibinfo {author} {\bibfnamefont {P.}~\bibnamefont {Piot}}, \emph {et~al.},\ }\bibfield  {title} {\bibinfo {title} {Demonstration of sub-gv/m accelerating field in a photoemission electron gun powered by nanosecond x-band radio-frequency pulses},\ }\href@noop {} {\bibfield  {journal} {\bibinfo  {journal} {Physical Review Accelerators and Beams}\ }\textbf {\bibinfo {volume} {25}},\ \bibinfo {pages} {083402} (\bibinfo {year} {2022})}\BibitemShut {NoStop}%
\bibitem [{\citenamefont {Chen}\ \emph {et~al.}(2025)\citenamefont {Chen}, \citenamefont {Doran}, \citenamefont {Liu}, \citenamefont {Piot}, \citenamefont {Power}, \citenamefont {Whiteford}, \citenamefont {Wisniewski}, \citenamefont {Jing}, \citenamefont {Knight},\ and\ \citenamefont {Kuzikov}}]{chen2025design}%
  \BibitemOpen
  \bibfield  {author} {\bibinfo {author} {\bibfnamefont {G.}~\bibnamefont {Chen}}, \bibinfo {author} {\bibfnamefont {S.}~\bibnamefont {Doran}}, \bibinfo {author} {\bibfnamefont {W.}~\bibnamefont {Liu}}, \bibinfo {author} {\bibfnamefont {P.}~\bibnamefont {Piot}}, \bibinfo {author} {\bibfnamefont {J.}~\bibnamefont {Power}}, \bibinfo {author} {\bibfnamefont {C.}~\bibnamefont {Whiteford}}, \bibinfo {author} {\bibfnamefont {E.}~\bibnamefont {Wisniewski}}, \bibinfo {author} {\bibfnamefont {C.}~\bibnamefont {Jing}}, \bibinfo {author} {\bibfnamefont {E.}~\bibnamefont {Knight}},\ and\ \bibinfo {author} {\bibfnamefont {S.}~\bibnamefont {Kuzikov}},\ }\bibfield  {title} {\bibinfo {title} {Design and photoemission studies of a high-gradient x-band photogun operating in the short-pulse regime},\ }\href@noop {} {\bibfield  {journal} {\bibinfo  {journal} {Nuclear Instruments and Methods in Physics Research Section A: Accelerators, Spectrometers, Detectors and Associated Equipment}\ ,\ \bibinfo {pages} {170205}} (\bibinfo
  {year} {2025})}\BibitemShut {NoStop}%
\bibitem [{\citenamefont {Shao}\ \emph {et~al.}(2022)\citenamefont {Shao}, \citenamefont {Chen}, \citenamefont {Doran}, \citenamefont {Ha}, \citenamefont {Jing}, \citenamefont {Lin}, \citenamefont {Liu}, \citenamefont {Peng}, \citenamefont {Power}, \citenamefont {Shi} \emph {et~al.}}]{shao2022demonstration}%
  \BibitemOpen
  \bibfield  {author} {\bibinfo {author} {\bibfnamefont {J.}~\bibnamefont {Shao}}, \bibinfo {author} {\bibfnamefont {H.}~\bibnamefont {Chen}}, \bibinfo {author} {\bibfnamefont {D.}~\bibnamefont {Doran}}, \bibinfo {author} {\bibfnamefont {G.}~\bibnamefont {Ha}}, \bibinfo {author} {\bibfnamefont {C.}~\bibnamefont {Jing}}, \bibinfo {author} {\bibfnamefont {X.}~\bibnamefont {Lin}}, \bibinfo {author} {\bibfnamefont {W.}~\bibnamefont {Liu}}, \bibinfo {author} {\bibfnamefont {M.}~\bibnamefont {Peng}}, \bibinfo {author} {\bibfnamefont {J.}~\bibnamefont {Power}}, \bibinfo {author} {\bibfnamefont {J.}~\bibnamefont {Shi}}, \emph {et~al.},\ }\bibfield  {title} {\bibinfo {title} {Demonstration of gradient above 300 mv/m in short pulse regime using an x-band single-cell structure},\ }\href@noop {} {\bibfield  {journal} {\bibinfo  {journal} {IPAC22, Bangkok, Thailand}\ } (\bibinfo {year} {2022})}\BibitemShut {NoStop}%
\bibitem [{\citenamefont {Hopkins}\ \emph {et~al.}(1984)\citenamefont {Hopkins}, \citenamefont {Sessler},\ and\ \citenamefont {Wurtele}}]{hopkin-1984-a}%
  \BibitemOpen
  \bibfield  {author} {\bibinfo {author} {\bibfnamefont {D.}~\bibnamefont {Hopkins}}, \bibinfo {author} {\bibfnamefont {A.}~\bibnamefont {Sessler}},\ and\ \bibinfo {author} {\bibfnamefont {J.}~\bibnamefont {Wurtele}},\ }\bibfield  {title} {\bibinfo {title} {The two-beam accelerator},\ }\href {https://doi.org/https://doi.org/10.1016/0168-9002(84)90004-4} {\bibfield  {journal} {\bibinfo  {journal} {Nuclear Instruments and Methods in Physics Research Section A: Accelerators, Spectrometers, Detectors and Associated Equipment}\ }\textbf {\bibinfo {volume} {228}},\ \bibinfo {pages} {15} (\bibinfo {year} {1984})}\BibitemShut {NoStop}%
\bibitem [{\citenamefont {Schnell}(1985)}]{schnell-1985-a}%
  \BibitemOpen
  \bibfield  {author} {\bibinfo {author} {\bibfnamefont {W.}~\bibnamefont {Schnell}},\ }\href {https://cds.cern.ch/record/255337} {\emph {\bibinfo {title} {{Consideration of a two-beam twin RF scheme for powering an RF linear collider}}}},\ \bibinfo {type} {Tech. Rep.}\ (\bibinfo  {institution} {CERN},\ \bibinfo {address} {Geneva},\ \bibinfo {year} {1985})\BibitemShut {NoStop}%
\bibitem [{\citenamefont {Sessler}\ and\ \citenamefont {Yu}(1987)}]{sessler:1987-a}%
  \BibitemOpen
  \bibfield  {author} {\bibinfo {author} {\bibfnamefont {A.~M.}\ \bibnamefont {Sessler}}\ and\ \bibinfo {author} {\bibfnamefont {S.~S.}\ \bibnamefont {Yu}},\ }\bibfield  {title} {\bibinfo {title} {Relativistic klystron two-beam accelerator},\ }\href {https://doi.org/10.1103/PhysRevLett.58.2439} {\bibfield  {journal} {\bibinfo  {journal} {Phys. Rev. Lett.}\ }\textbf {\bibinfo {volume} {58}},\ \bibinfo {pages} {2439} (\bibinfo {year} {1987})}\BibitemShut {NoStop}%
\bibitem [{\citenamefont {Corsini}(2017)}]{corsini-2017-a}%
  \BibitemOpen
  \bibfield  {author} {\bibinfo {author} {\bibfnamefont {R.}~\bibnamefont {Corsini}},\ }\bibfield  {title} {\bibinfo {title} {{F}inal {R}esults {F}rom the {C}lic {T}est {F}acility ({CTF}3)},\ }in\ \href {https://doi.org/10.18429/JACoW-IPAC2017-TUZB1} {\emph {\bibinfo {booktitle} {Proc. of International Particle Accelerator Conference (IPAC'17), Copenhagen, Denmark, 14-19 May, 2017}}},\ \bibinfo {series and number} {\bibinfo {series} {International Particle Accelerator Conference}\ No.~\bibinfo {number} {8}}\ (\bibinfo  {publisher} {JACoW},\ \bibinfo {address} {Geneva, Switzerland},\ \bibinfo {year} {2017})\ pp.\ \bibinfo {pages} {1269--1274},\ \bibinfo {note} {https://doi.org/10.18429/JACoW-IPAC2017-TUZB1}\BibitemShut {NoStop}%
\bibitem [{\citenamefont {Jing}\ \emph {et~al.}(2012)\citenamefont {Jing}, \citenamefont {Antipov}, \citenamefont {Schoessow},\ and\ \citenamefont {Kanareykin}}]{jing2012high}%
  \BibitemOpen
  \bibfield  {author} {\bibinfo {author} {\bibfnamefont {C.}~\bibnamefont {Jing}}, \bibinfo {author} {\bibfnamefont {S.}~\bibnamefont {Antipov}}, \bibinfo {author} {\bibfnamefont {P.}~\bibnamefont {Schoessow}},\ and\ \bibinfo {author} {\bibfnamefont {A.}~\bibnamefont {Kanareykin}},\ }\bibfield  {title} {\bibinfo {title} {High frequency high power rf generation using a relativistic electron beam},\ }\href@noop {} {\bibfield  {journal} {\bibinfo  {journal} {IPAC2012-ProceedingsNew Orleans, Louisiana, USA}\ } (\bibinfo {year} {2012})}\BibitemShut {NoStop}%
\bibitem [{\citenamefont {Shi}\ \emph {et~al.}(2013)\citenamefont {Shi}, \citenamefont {Wu}, \citenamefont {Jing}, \citenamefont {Yang}, \citenamefont {Zha}, \citenamefont {Gao}, \citenamefont {Gai},\ and\ \citenamefont {Chen}}]{shi2013development}%
  \BibitemOpen
  \bibfield  {author} {\bibinfo {author} {\bibfnamefont {J.}~\bibnamefont {Shi}}, \bibinfo {author} {\bibfnamefont {X.}~\bibnamefont {Wu}}, \bibinfo {author} {\bibfnamefont {C.}~\bibnamefont {Jing}}, \bibinfo {author} {\bibfnamefont {Y.}~\bibnamefont {Yang}}, \bibinfo {author} {\bibfnamefont {H.}~\bibnamefont {Zha}}, \bibinfo {author} {\bibfnamefont {Q.}~\bibnamefont {Gao}}, \bibinfo {author} {\bibfnamefont {W.}~\bibnamefont {Gai}},\ and\ \bibinfo {author} {\bibfnamefont {H.}~\bibnamefont {Chen}},\ }\bibfield  {title} {\bibinfo {title} {Development of an x-band metallic power extractor for the argonne wakefield accelerator},\ }in\ \href@noop {} {\emph {\bibinfo {booktitle} {Proc. 4th Int. Particle Accelerator Conf.(IPAC’13)}}}\ (\bibinfo {year} {2013})\ pp.\ \bibinfo {pages} {2771--2773}\BibitemShut {NoStop}%
\bibitem [{\citenamefont {Jing}\ \emph {et~al.}(2018)\citenamefont {Jing}, \citenamefont {Antipov}, \citenamefont {Conde}, \citenamefont {Gai}, \citenamefont {Ha}, \citenamefont {Liu}, \citenamefont {Neveu}, \citenamefont {Power}, \citenamefont {Qiu}, \citenamefont {Shi} \emph {et~al.}}]{jing2018electron}%
  \BibitemOpen
  \bibfield  {author} {\bibinfo {author} {\bibfnamefont {C.}~\bibnamefont {Jing}}, \bibinfo {author} {\bibfnamefont {S.}~\bibnamefont {Antipov}}, \bibinfo {author} {\bibfnamefont {M.}~\bibnamefont {Conde}}, \bibinfo {author} {\bibfnamefont {W.}~\bibnamefont {Gai}}, \bibinfo {author} {\bibfnamefont {G.}~\bibnamefont {Ha}}, \bibinfo {author} {\bibfnamefont {W.}~\bibnamefont {Liu}}, \bibinfo {author} {\bibfnamefont {N.}~\bibnamefont {Neveu}}, \bibinfo {author} {\bibfnamefont {J.}~\bibnamefont {Power}}, \bibinfo {author} {\bibfnamefont {J.}~\bibnamefont {Qiu}}, \bibinfo {author} {\bibfnamefont {J.}~\bibnamefont {Shi}}, \emph {et~al.},\ }\bibfield  {title} {\bibinfo {title} {Electron acceleration through two successive electron beam driven wakefield acceleration stages},\ }\href@noop {} {\bibfield  {journal} {\bibinfo  {journal} {Nuclear Instruments and Methods in Physics Research Section A: Accelerators, Spectrometers, Detectors and Associated Equipment}\ }\textbf {\bibinfo {volume} {898}},\ \bibinfo {pages} {72}
  (\bibinfo {year} {2018})}\BibitemShut {NoStop}%
\bibitem [{CST(2021)}]{CST}%
  \BibitemOpen
  \href {https://www.3ds.com/products/simulia/cst-studio-suite/} {\bibinfo {title} {{\sc CST Studio Suite} electromagnetic field simulation software}} (\bibinfo {year} {2021})\BibitemShut {NoStop}%
\bibitem [{\citenamefont {Peng}\ \emph {et~al.}(2019)\citenamefont {Peng}, \citenamefont {Shao}, \citenamefont {Jing}, \citenamefont {Wisniewski}, \citenamefont {Ha}, \citenamefont {Seok}, \citenamefont {Conde},\ and\ \citenamefont {Liu}}]{peng2019generation}%
  \BibitemOpen
  \bibfield  {author} {\bibinfo {author} {\bibfnamefont {M.}~\bibnamefont {Peng}}, \bibinfo {author} {\bibfnamefont {J.}~\bibnamefont {Shao}}, \bibinfo {author} {\bibfnamefont {C.}~\bibnamefont {Jing}}, \bibinfo {author} {\bibfnamefont {E.}~\bibnamefont {Wisniewski}}, \bibinfo {author} {\bibfnamefont {G.}~\bibnamefont {Ha}}, \bibinfo {author} {\bibfnamefont {J.}~\bibnamefont {Seok}}, \bibinfo {author} {\bibfnamefont {M.}~\bibnamefont {Conde}},\ and\ \bibinfo {author} {\bibfnamefont {W.}~\bibnamefont {Liu}},\ }\bibfield  {title} {\bibinfo {title} {Generation of high power short rf pulses using an x-band metallic power extractor driven by high charge multi-bunch train},\ }in\ \href@noop {} {\emph {\bibinfo {booktitle} {Proc. 10th International Particle Accelerator Conference (IPAC’19), Melbourne, Australia}}}\ (\bibinfo {year} {2019})\ pp.\ \bibinfo {pages} {734--737}\BibitemShut {NoStop}%
\bibitem [{\citenamefont {Carlsten}(1989)}]{CARLSTEN1989313}%
  \BibitemOpen
  \bibfield  {author} {\bibinfo {author} {\bibfnamefont {B.}~\bibnamefont {Carlsten}},\ }\bibfield  {title} {\bibinfo {title} {New photoelectric injector design for the los alamos national laboratory xuv fel accelerator},\ }\href {https://doi.org/https://doi.org/10.1016/0168-9002(89)90472-5} {\bibfield  {journal} {\bibinfo  {journal} {Nuclear Instruments and Methods in Physics Research Section A: Accelerators, Spectrometers, Detectors and Associated Equipment}\ }\textbf {\bibinfo {volume} {285}},\ \bibinfo {pages} {313} (\bibinfo {year} {1989})}\BibitemShut {NoStop}%
\end{thebibliography}

%

\end{document}